\documentclass[preprint,journal]{vgtc}       

\ifpdf
  \pdfoutput=1\relax                   
  \usepackage{graphicx}                
  \DeclareGraphicsExtensions{.pdf,.png,.jpg,.jpeg} 
\else
  \usepackage{graphicx}                
  \DeclareGraphicsExtensions{.eps}     
\fi%

\graphicspath{{figures/}{pictures/}{images/}{./}} 

\usepackage{microtype}                 
\PassOptionsToPackage{warn}{textcomp}  
\usepackage{textcomp}                  
\usepackage{mathptmx}                  
\usepackage{times}                     
\usepackage{cite}                      
\usepackage{tabu}                      
\usepackage{booktabs}                  

\usepackage{paralist}
\usepackage{xcolor}
\usepackage[normalem]{ulem}
\usepackage{mathtools}
\usepackage{amssymb}
\usepackage{hyperref}
\usepackage{multirow}

\usepackage{graphicx}
\usepackage{array}
\usepackage{balance}

\usepackage[detect-none]{siunitx}
\useunder{\uline}{\ul}{}
\newcommand{\name}{{\textit{DeepDrawing}}}

\newcommand{\newtext}[1]{\textcolor{black}{#1}}

\ieeedoi{10.1109/TVCG.2019.2934798}

\onlineid{1046}

\vgtccategory{Research}
\vgtcpapertype{algorithm/technique}

\title{\name: A Deep Learning Approach to Graph Drawing}

\author{Yong Wang, Zhihua Jin, Qianwen Wang, Weiwei Cui, Tengfei Ma and Huamin Qu}
\authorfooter{
\item
 Y. Wang, Z. Jin, Q. Wang and H. Qu are with the Hong Kong University of Science and Technology (HKUST). E-mail: \{ywangct, zjinak, qwangbb, huamin\}@cse.ust.hk. Z. Jin is also affiliated with Zhejiang University.
 
\item
 W. Cui is with Microsoft Research Asia. E-mail: weiwei.cui@microsoft.com.
\item
 T. Ma is with IBM T. J. Watson Research Center. E-mail: tengfei.ma1@ibm.com.
}

\shortauthortitle{Biv \MakeLowercase{\textit{et al.}}: Global Illumination for Fun and Profit}
\abstract{Node-link diagrams are widely used to facilitate network explorations. However, when using a graph drawing technique to visualize networks, users often need to tune different algorithm-specific parameters iteratively by comparing the corresponding drawing results in order to achieve a desired visual effect. This trial and error process is often tedious and time-consuming, especially for non-expert users. 
\newtext{
Inspired by the powerful data modelling and prediction capabilities 
of deep learning techniques, we explore the possibility of applying deep learning techniques to graph drawing. Specifically, we propose using a graph-LSTM-based approach to directly map network structures to graph drawings. Given a set of layout examples as the training dataset, we train the proposed graph-LSTM-based model to capture their layout characteristics. 
Then, the trained model is used to generate graph drawings in a similar style for new networks.
We evaluated the proposed approach on two special types of layouts (i.e., grid layouts and star layouts) and two general types of layouts (i.e., ForceAtlas2 and PivotMDS) in both qualitative and quantitative ways. The results provide support for the effectiveness of our approach.
We also conducted a time cost assessment on the drawings of small graphs with 20 to 50 nodes. We further report the lessons we learned and discuss the limitations and future work.
}
}


\CCScatlist{ 
 \CCScat{K.6.1}{Management of Computing and Information Systems}%
{Project and People Management}{Life Cycle};
 \CCScat{K.7.m}{The Computing Profession}{Miscellaneous}{Ethics}
}

\vgtcinsertpkg

\begin{document}

\maketitle


\section{Introduction}
\label{sec_intro}

Node-link diagrams are widely used to visualize networks in various areas, such as bioinformatics, finance, and social networks analysis. Many graph drawing techniques have been proposed in the past five decades~\cite{battista1998graph,herman2000graph,gibson2013survey,kaufmann2003drawing} to achieve desired visual properties of node-link diagrams, such as fewer edge crossings, less node occlusion, and better community preservation, to support an easy interpretation of the underlying network structures.
\newtext{
Graph drawing methods are often based on different underlying principles: from spring-embedder algorithms~\cite{frick1994fast}, to energy-based approaches~\cite{kamada1989algorithm,noack2007energy,jacomy2014forceatlas2}, to dimension-reduction based techniques~\cite{brandes2006eigensolver,gansner2004graph}.
When users employ a specific graph drawing algorithm, they usually need to understand its basic mechanism and tune its various parameters to achieve the desired visual properties for different graphs, though some default parameters are often provided by the developers.
Such trial-and-error process requires time and is a non-trivial challenge for less experienced users without a background in graph drawing.
Since the algorithm-specific parameters and the corresponding drawings often depend on the input graph structure, we consider the question whether a machine learning approach can be used instead to generate the graph drawings.
}

\newtext{One possible choice is using the graph structure information to directly predict the graph drawings with certain visual properties, where graph drawing is considered as a structure-to-layout mapping function.}
A recent work~\cite{kwon2018would} speeds up the graph drawing process by showing a pre-computed drawing of a graph that has a similar graphlet frequency to the input graph.
However, such an approximation needs to extract a hand-crafted feature (i.e., graphlet frequency) and cannot guarantee that it will definitely generate an accurate drawing for the input graph, since graphs with a similar graphlet frequency can have totally-different topology structures. 
On the other hand, deep learning techniques have shown a powerful capability for modelling the training data and making predictions for relevant data inputs, where no hand-crafted features are needed. Deep learning techniques have been successfully used in various applications such as computer vision and natural language processing fields~\cite{goodfellow2016deep,lecun2015deep}.
\newtext{
Inspired by these successes, we are exploring the possibility of applying deep learning techniques to the problem of graph drawing in this paper.
}


However, we are not aware of any prior work on using deep learning for graph drawing, and there is still a significant gap in how to achieve this. This gap is mainly due to three aspects: model architecture, loss function design, and training data.


\textbf{Model architecture:} Graphs represent topological relationships between different entities; this makes graphs intrinsically different from typical datasets that are often used in deep learning, such as images, videos and texts, which are all Euclidean data.
Therefore, it remains unclear whether deep learning techniques can be used for graph drawing and how to adapt existing techniques for the graph drawing problem.
Some recent work on Graph Convolutional Neural Networks (GCN)~\cite{kipf2016semi,defferrard2016gcn} has adapted the CNN framework for graph data, but they are mainly applied to node classification and link prediction tasks,
\newtext{which is different from ours.}
Since there is no prior work on this problem, our first step is identifying a deep neural network model that can be used to predict graph drawings. At the same time, it is also necessary to identify a transformation that can convert a graph structure into a data structure that can be processed by deep learning models.
%
%
%
%

\textbf{Loss function design:} One key part of using deep learning techniques is designing an appropriate loss function to guide the model training.
For typical deep learning tasks such as classification, the loss function can be easily defined by counting incorrect predictions. However, it is much more complicated for graph drawing. For example, how can we define whether the prediction of a node position is ``correct'' or ``incorrect''? Since a graph drawing may have significantly-different visual appearances after linear transformations, like translation, rotation and scaling, it is also critical for us to design a loss function that is invariant to those transformations.

\textbf{Training data:}
High-quality training datasets are critical for using deep learning techniques and many benchmark datasets have been published in different applications, e.g., ImageNet dataset~\footnote{\href{http://www.image-net.org/}{\url{http://www.image-net.org/}}} for image classification, MNIST dataset~\footnote{\href{http://yann.lecun.com/exdb/mnist/}{\url{http://yann.lecun.com/exdb/mnist/}}} for digits recognition.
However, there are no available benchmark datasets with clear drawing labels for graph drawing tasks.



In this paper, we propose a graph-LSTM-based approach to directly generate graph drawing results based on the topology structures of input graphs. We transform the graph topology information into a sequence of adjacency vectors using Breadth First Search (BFS), where each adjacency vector encodes the connection information between each node and its adjacent nodes in the sequence. 
\newtext{In addition, we propose a Procrustes Statistics based loss function, which is essentially invariant to translation, rotation and scaling of graph drawing, to assess the learning quality and guide the model training.
Furthermore, we generate three graph datasets (including grid graphs, star graphs and general graphs with clear communities), which are further drawn by both two regular drawings (i.e., grid layout and star layout) and two general drawings 
(i.e., a force-directed graph drawing~\cite{jacomy2014forceatlas2} and a dimension-reduction-based drawing~\cite{brandes2006eigensolver}).
We carefully choose the parameters of drawing algorithms to
generate drawing results with certain desired visual properties for these graphs (e.g., better preservation of community structure and less node occlusion).
As a proof of concept, these drawings are treated as the ground-truth labels.
The graphs and the corresponding drawings are used to train and test the proposed
approach.
}

\newtext{
We investigated the effectiveness of the proposed deep learning approach through both qualitative comparisons and quantitative metric evaluations,
where the drawings of our approach are compared with the ground truth drawings (drawn by ForceAtlas2~\cite{jacomy2014forceatlas2}
and PivotMDS~\cite{brandes2006eigensolver}) and the drawings by a 4-layer bidirectional LSTM model.}
In summary, the primary contributions of this work include: 
\begin{compactitem}

\item A novel graph-LSTM-based approach for graph drawing, which, to the best of our knowledge, is the first time that deep learning has been applied to graph drawing.

\item \newtext{Qualitative and quantitative evaluations on three synthetic graph datasets (i.e., grid graphs, star graphs and clustered graphs with \numrange{20}{50} nodes) and four types of drawings (i.e., grid layout, star layout, ForceAtlas2 and PivotMDS), which provides support for the effectiveness and efficiency of our approach in generating graph drawings similar to the training data.}

\item \newtext{A detailed summary of the lessons we learned in the development of the proposed approach}, which, we hope, will assist in future research on using deep learning for graph visualization.

\end{compactitem}


\section{Related Work}
This section summarizes the related work of this paper, which mainly consists of three parts: graph drawing, graph neural networks, and machine learning approaches to graph drawing.

\subsection{Graph Drawing} 
\label{sec_graph_drawing}
\newtext{
One of the central problems in graph visualization is the design of the algorithms for graph layout.}
Since Tutte~\cite{tutte1963draw,tutte1960convex} proposed his barycenter method for graph drawing more than fifty years ago, the information visualization community has proposed many graph drawing techniques. These algorithms can be found in various books~\cite{battista1998graph,kaufmann2003drawing,tamassia2013handbook} and surveys~\cite{herman2000graph,di1994algorithms,gibson2013survey,von2011visual,yoghourdjian2018exploring}. 

\newtext{
Typically, graph drawing algorithms generate only one drawing for a graph, though some work~\cite{biedl1998graph} also proposes producing multiple drawings for the same graph.}
According to the survey by Gibson~\textit{et al.}~\cite{gibson2013survey}, the existing graph drawing algorithms can be categorized into three types: force-directed layouts, dimension reduction based layouts, and computational improvements like multi-level techniques. Force-directed graph layout approaches 
regard a graph as a physical system, where nodes are attracted and repelled in order to achieve desirable graph drawing aesthetics. Eades~\cite{eades1984heuristic} proposed a spring-electrical-based graph drawing approach, where nodes and edges are modeled as steel rings and springs, respectively. The final graph drawing result is the stable state when the forces on each node reach an equilibrium. This kind of modelling is the start of all force-directed techniques and has inspired many follow-up algorithms, like the spring-embedder algorithm by Fruchterman and Reingold~\cite{fruchterman1991graph}, the graph-embedder (GEM) algorithm~\cite{frick1994fast}, and the energy-based approaches~\cite{kamada1989algorithm,noack2007energy,jacomy2014forceatlas2,wang2017revisiting}.
Dimension reduction based methods focus on retaining the information of high-dimensional space in the projected 2D plane, especially the graph-theoretic distance between a pair of nodes. Various dimension reduction techniques have been used for graph drawing, including multidimensional scaling (MDS)~\cite{brandes2006eigensolver,gansner2004graph}, linear dimension reduction~\cite{harel2002graph}, self-organising maps (SOM)~\cite{bonabeau2002graph,bonabeau1998self} and t-SNE~\cite{kruiger2017graph}. 
The last category of algorithms mainly aims to improve the efficiency of force-directed algorithms for drawing very large graphs. These approaches often follow a multi-level paradigm: optimizing the graph drawing in a coarser graph representation and further propagate the layout result back to the original graph~\cite{hu2005efficient,hachul2004drawing,harel2000fast,gajer2000multi}. 

\newtext{
Different from prior studies, 
this paper explores the possibility of using deep neural networks for graph drawing.}






\subsection{Graph Neural Networks}
\label{sec_gnn}
Existing deep neural networks mainly focus on regular Euclidean data (e.g., images and text), which cannot be directly applied to non-Euclidean data, like graphs. 
To address this issue, a number of graph neural networks (GNN) have been proposed by extending existing deep neural networks, e.g., convolutional neural networks (CNNs) and recurrent neural networks (RNNs), to the graph domain~\cite{zhou2018gnnsurvey}.

The GNNs that are derived from CNNs can be categorized into spectral approaches and non-spectral approaches~\cite{zhou2018gnnsurvey}.
Spectral approaches apply convolution to the spectral representation
of graphs~\cite{henaff2015deep,kipf2016semi,bruna2014spectral, brandes2006eigensolver, defferrard2016gcn}.
For example, Bruna et al.~\cite{bruna2014spectral} conducted convolution in the Fourier domain using eigen decomposition of the graph Laplacian.
Defferrard et al.~\cite{defferrard2016gcn} approximated the spectral convolution using Chebyshev polynomials and reduced the computational cost.
Since spectral convolution depends on the input graph, spectral approaches are usually applied in the learning problem within a single graph.
Instead of defining convolution operations in the spectral field,
non-spectral approaches operate convolution directly on the graph 
~\cite{hamilton2017inductive, monti2017geometric, niepert2016learning}.
The key challenge of non-spectral approaches is how to define the neighborhood of a node as the receptive field and various methods have been proposed, including adaptive weight matrices~\cite{duvenaud2015convolutional}, uniformly sampling~\cite{hamilton2017inductive}, and transition matrices~\cite{atwood2016diffusion}.
%
A closely related research direction explores using RNNs for graph-structured data~\cite{zhou2018gnnsurvey, peng2017graphlstm, tai2015tree-lstm, li2015gatedgraph}.
For example, Li et al.~\cite{li2015gatedgraph} modified the Gate Recurrent Units (GRU) and proposed a gated GNN to learn node representations.
Tai et al.~\cite{tai2015tree-lstm} proposed two types of tree-LSTM, generalizing the basic LSTM to tree-structure typologies, \newtext{to predict the semantic relatedness of sentences}.
Peng et al. ~\cite{peng2017graphlstm} extended tree-LSTM by distinguishing different edge types in the graph and applied the model to the relation extraction problem in the Natural Language Processing (NLP) field.
You et al.~\cite{you2018graphrnn} developed an RNN-based method for modeling complex distributions over multiple graphs and further generating graphs.

\newtext{
However, the idea of applying GNNs to graph drawing has been rarely explored, even though it is a fundamental research direction in the visualization community.}

\subsection{Machine Learning Approaches to Graph Drawing}
According to the survey by Santos Vieira~\textit{et al.}~\cite{dos2015application}, there have been only a few studies about applying machine learning techniques to graph drawing. These techniques can be roughly classified into two categories: the approaches that learn from human interaction and those without using human interaction.
The first group of techniques assume that the choices of aesthetic criteria and their importance depend on the users' subjective preferences. Therefore, these approaches keep humans in the loop and use evolutionary algorithms (e.g., genetic algorithms) to learn user preferences~\cite{masui1994evolutionary,rosete1999study,barbosa2001interactive,bach2012interactive,sponemann2014evolutionary}. However, these approaches are inherently dependant on user interactions.
The second category focuses on using traditional neural-network-based algorithms to optimize the aesthetic criteria of a graph layout~\cite{cimikowski1996neural,wang2005artificial} or to draw graphs in both 2D and 3D space~\cite{meyer1998self}. 
However, these early studies are essentially categorized as traditional graph drawing methods, where algorithm-specific parameters are still needed.



Recently, Kwon~\textit{et al.}~\cite{kwon2018would} proposed a machine learning approach that provides users with a quick preview of the graph drawing and it uses graphlet frequency to compute the similarities among different graph structures. However, as Kwon~\textit{et al.} pointed out in their paper, similar graphlet frequencies do not necessarily lead to similar drawings.
\newtext{Deep learning techniques have recently been applied to multidimensional projections~\cite{espadoto2019deep}. However, the networks cannot be directly used for graph drawing, since the designs of the model input and training loss function for graph drawing are significantly different from those of multidimensional projection.
Also, neural-network-based approaches have been proposed to evaluate graph drawing results~\cite{klammler2018aesthetic,haleem2018evaluating}.
}

This paper expands on these earlier findings and makes new contributions to the field by focusing on applying deep learning techniques to direct graph drawing instead of using a similar preview, and the trained model can be reusable for new graphs.

\begin{figure}[]
\centering 
\includegraphics[width=0.75\columnwidth]{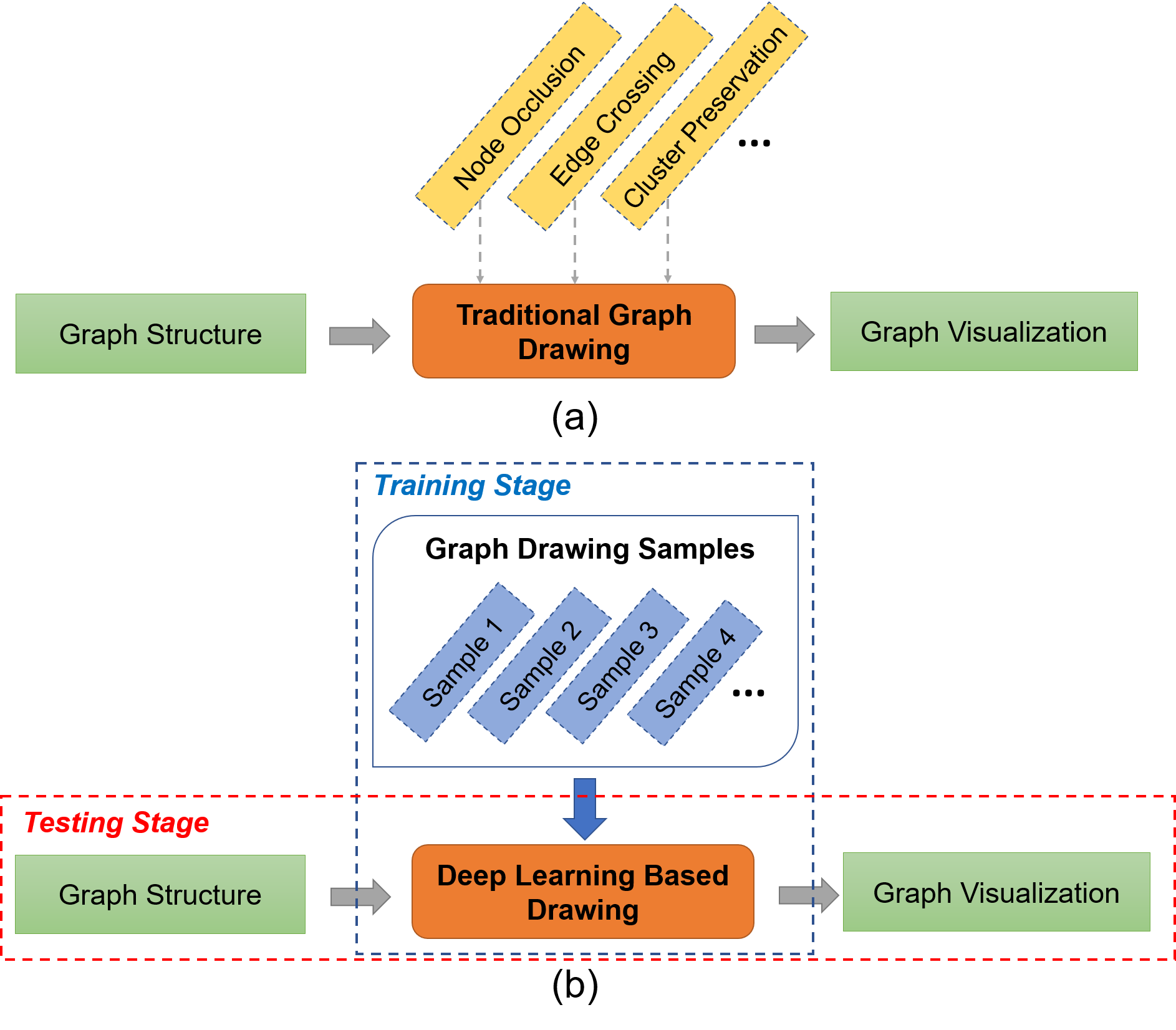}
\vspace{-4mm}
\caption{The workflow of graph drawing algorithms: (a) traditional graph drawing algorithms, (b) the proposed deep learning based approach.}
\label{fig_workflow}
\vspace{-3mm}
\end{figure}


\section{Problem Statement}





A graph $G = (V, E)$ consists of a set of $n$ nodes $V = \{v_1, v_2, ..., v_n\}$ and an edge set $E \subseteq V \times V$. 
The graphs can be classified into directed and un-directed graphs, depending on whether the edges are directed or not.
For graph drawing, the edge directions are often ignored, since they can easily be visualized by adding an arrow to each link.
In this paper, we focus on the visualization of unweighted and undirected graphs.
\newtext{The graph drawing problem is finding a set of coordinates $C=\{c_v|v\in V\}$ for the given graph $G=\{V,E\}$~\cite{hu2005efficient}.}
In this paper, we only consider 2D drawings, which means $p_v\in R^2$.
\newtext{Also, we assume that the edges in the graph drawings are straight-lines, instead of arcs or curves.}

As discussed in Section~\ref{sec_graph_drawing}, there have been many graph drawing algorithms that are proposed for optimizing aesthetic criteria like minimizing edge crossings, avoiding node occlusions, and preserving the node community structures. These criteria are formulated as objective functions and integrated into the design of \textit{traditional graph drawing algorithms} (Fig.~\ref{fig_workflow}(a)). 
When using a drawing algorithm to visualize a specific graph, users also need to tune the algorithm parameters through trial and error to achieve a suitable graph drawing result.


In this paper, we formalize graph drawing as a learning problem and propose a novel and generalizable deep-learning-based approach to graph drawing (Fig.~\ref{fig_workflow}(b)).
Given a set of graph drawing examples with desirable aesthetic properties and their structures, the deep learning model is trained to learn the mapping and corresponding algorithm-specific parameters for determining the desirable graph drawings (\textit{Training Stage} in Fig.~\ref{fig_workflow}(b)).
Once the deep learning model is successfully trained, when given a new graph, it can automatically analyze the graph structure and directly generate a layout that carries
\newtext{the common visual properties of the drawing examples}
(\textit{Testing Stage} in Fig.~\ref{fig_workflow}(b)).
\newtext{We use the term \emph{``graph drawing style"} to refer to the common visual properties (e.g., the characteristics regarding edge crossings, community preservation and node occlusion) that are shared by the training graph drawings. The deep learning model learns one specific drawing style from a certain training dataset.}
\newtext{In real applications, the deep learning model can be pre-trained by experts. Then, the well-trained model can be directly used by different users to visualize graphs, especially graph drawing novices.}

\section{Background: Long Short-Term Memory Networks}
\label{sec_lstm}


As will be introduced in Section~\ref{sec_method}, we propose a graph-LSTM-based approach for graph drawing, where the foundation of our model is the typical LSTM architecture. This section introduces the basic concepts and other related background of LSTM.

LSTM architecture is a popular variant of Recurrent Neural Networks (RNNs). 
It can learn long-distance dependencies of the input sequence and can avoid the gradient exploding and vanishing problems of traditional RNN models~\cite{hochreiter1997long,gers1999learning}. The main reason for this is that LSTM models introduce a \emph{memory cell}, which can preserve the state over a long time range. An LSTM memory cell consists of an input gate, output gate, and forget gate.
It takes sequential data $\{x_0, ..., x_T\}$ as inputs and maintains a time-variant \emph{cell state} vector $c_t$ and \emph{hidden state} vector $h_t$, where $x_t \in \mathbb{R}^m$, $c_t \in \mathbb{R}^n$ and $h_t \in \mathbb{R}^n$.

The transitions functions of LSTM are as follows: 
\vspace{-2mm}
\begin{align}\label{eq_lstm}
    i _ { t }  &= \sigma \left( W^{(i)} x_{ t } + U^{(i)}h_{t-1} + b^{(i)} \right) \\
    o _ { t }  &= \sigma \left( W^{(o)} x _ { t } + U^{(o)}h_{t-1} + b^{(o)} \right)\\
    \tilde { c }_{ t }  &= \tanh \left( W^{(c) } x _ { t } + U^{(c)}h_{t-1} + b^{(c)} \right)\\
    f_{t} &= \sigma \left( W^{(f)} x_{t} + U^{(f)} h_{t-1} + b^{(f)} \right)\\
    c_{t}  &= i_{ t } \odot \tilde {c}_{t} +  f _ { t } \odot c_{t-1}\\
    h_{t}  &= o_{ t } \odot \tanh \left( c_{ t } \right)  
\end{align} 

where $x_t$, $h_t$ and $c_{t} $ represent the input feature vector, hidden state, and cell state of the current time step $t$, respectively.
$W$'s and $U$'s are the weighted matrices for the input and hidden state, and $b$'s are the bias vectors. 
$\sigma$, $\tanh$ and $\odot$ are the sigmoid function, the hyperbolic tangent function, and the pointwise multiplication operation, respectively. 
Basically, the input gate $i_t$ controls how much the information is updated at each time step; the output gate $o_t$ controls how much of the internal state information flows out of the cell. The forget gate $f_{t}$ is a key design of LSTM; it enables LSTM models to forget the previous cell state that has become irrelevant to a certain degree.
Due to the design of these gates, LSTM models can learn and represent long-distance correlations within sequential input data~\cite{hochreiter1997long}.

\section{\name}
\label{sec_method}

We propose a deep learning based approach, called  \name, for graph drawing. Our model is built on the widely-used LSTM framework~\cite{hochreiter1997long}.
This section introduces {\name} from the perspectives of model architecture, input design, and loss function.


\subsection{Graph-LSTM-based Architecture}

When applying deep learning techniques to graph drawing, a fundamental requirement is to learn
\newtext{a certain graph drawing style}
from multiple graphs of various sizes.
As discussed in Section~\ref{sec_gnn}, many graph neural networks, like spectral approaches~\cite{henaff2015deep,bruna2014spectral,brandes2006eigensolver,defferrard2016gcn}, mainly focus on learning from a single graph or fixed-size graphs.
\newtext{Thus, these models do not easily generalize graphs with different sizes and structures~\cite{bresson2018experimental}.}
On the other hand, RNN-based graph neural networks are intrinsically applicable to graphs with variable sizes, since RNN cells can be recurrently used. Also, a recent study~\cite{you2018graphrnn} has shown that RNNs are capable of modelling the structure information of multiple graphs.
Inspired by these models, we focus on RNN-based approaches for graph drawing in this paper.

Among the RNN-based approaches, vanilla RNNs have proved to be difficult to train, as they suffer from gradient vanishing or explosion~\cite{bengio1994learning,pascanu2013difficulty} problems. 
On the contrary, as introduced in Section~\ref{sec_lstm}, LSTM models introduce a series of gates to avoid amplifying or suppressing the gradients, making them better at capturing long-distance dependencies.


In this paper, we propose a graph-LSTM-based approach for graph drawing.
The commonly-used LSTM architectures are often linearly chained, as described in Section~\ref{sec_lstm}. 
One of their major limitations is that they can only explicitly model sequential data. 
However, for graph drawing, the input is essentially the graph/network structure, which is usually not linearly-chained.
The layout position of a node in the graph drawing depends on
all the other nodes that are directly or indirectly connected to it. When using a general LSTM model, such kind of dependency information can still be weakened or lost, especially for the LSTM cells that are far from each other.
Inspired by the recent work in natural language processing field~\cite{tai2015tree-lstm,peng2017graphlstm},
\newtext{we propose adding direct connections between different LSTM cells to explicitly model the topological structure of input networks. Such direct connections are termed ``skip connections'' in the deep learning field~\cite{goodfellow2016deep}.}
Then we can use the linear chain between adjacent LSTM cells to propagate the overall state of prior graph nodes to the subsequent nodes along the chain.

\begin{figure}[h]
\centering 
\includegraphics[width=0.85\linewidth]{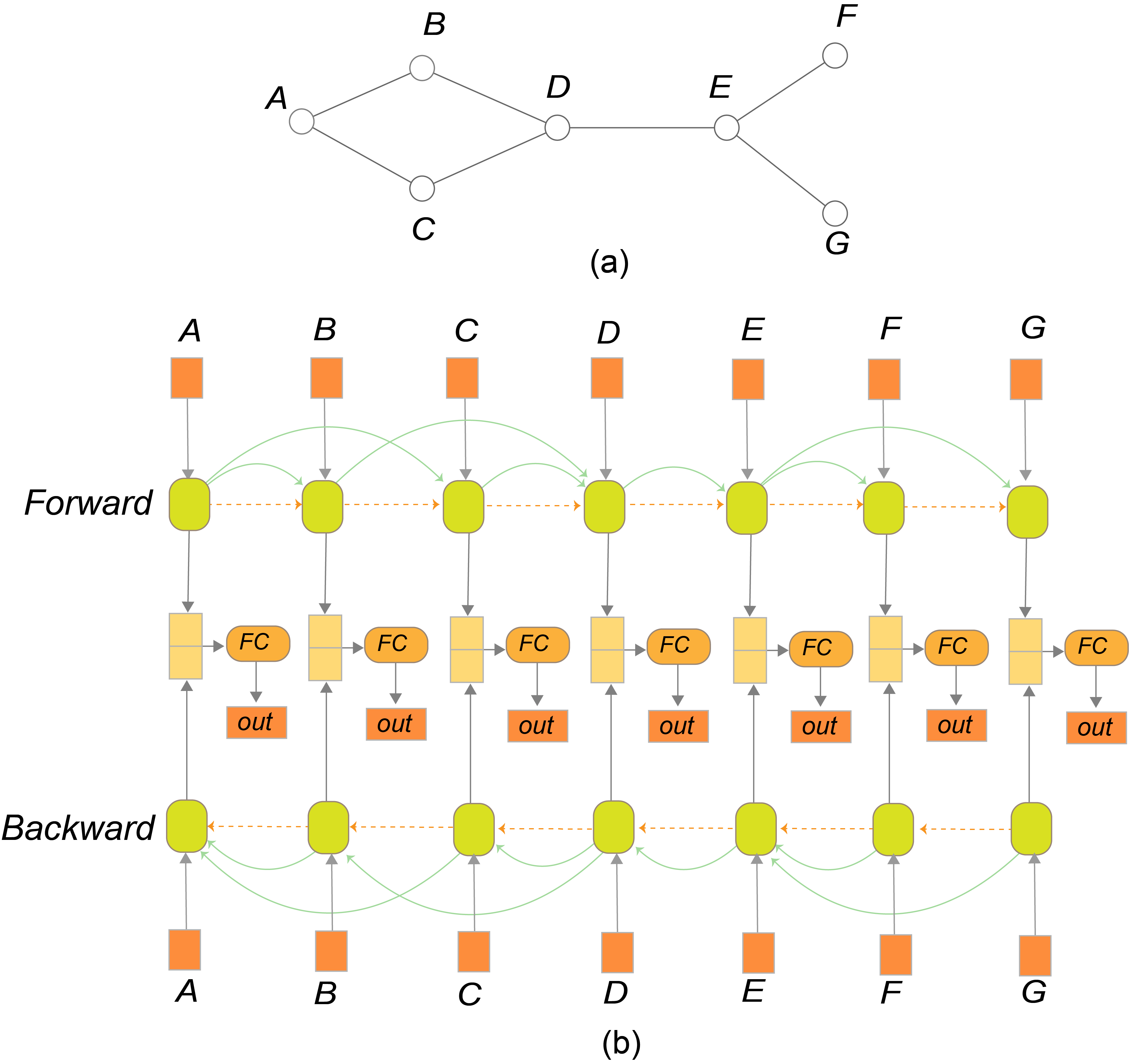}
\vspace{-3mm}
\caption{An illustration of the proposed graph-LSTM-based model architecture: (a) an example graph input, (b) the proposed model to process the input graph. The graph nodes are sorted using BFS, with each node represented by an adjacency vector encoding its connections with predecessor nodes.
The dotted yellow arrows (``fake'' edges) propagate the prior nodes' overall influence on the drawing of subsequent nodes, and the curved green arrows (real edges of graphs) explicitly reflect the actual graph structure, enhancing the graph drawing details.
The information of both forward and backward rounds is considered for generating the final 2D node layouts.
}
\label{fig_model}
\vspace{-3mm}
\end{figure}

\newtext{Although our model architecture is similar to prior studies~\cite{tai2015tree-lstm,peng2017graphlstm}, it targets at different problems. Unlike natural language processing (NLP) problems, where the input text is already sequential data and the input feature vector can be directly gained through word embedding, it is necessary to carefully design the architecture and input feature vector to model the graph topology information when using deep learning for graph drawing.}
To the best of our knowledge, our model is the first deep learning architecture proposed for graph drawing tasks.
Fig.~\ref{fig_model} provides an overview of the proposed model architecture. The input graph is transformed into a sequence of nodes. Each LSTM cell takes the feature vector of one node as input and generates the output status of each node. The green arrows between LSTM cells represent the \textit{\textbf{real edges}} in the graph structure, while the dotted yellow arrows are the \textbf{\textit{``fake'' edges}} between 
\newtext{adjacent nodes in the BFS-ordered node sequence}
to propagate the summary state of previous nodes to subsequent unprocessed nodes. The detailed transition equations of our model are as follows:

\vspace{-2mm}
\begin{align}\label{eq_graphlstm}
    i _ { t }  &= \sigma \left( W^{(i)} x_{ t } + U^{(i)}h_{t-1} + \sum _ { k \in P(t)} \tilde{U} ^ {(i)} h _ {k} + b^{(i)} \right) \\
    o _ { t }  &= \sigma \left( W^{(o)} x _ { t } + U^{(o)}h_{t-1} + \sum _ {k\in P(t)} \tilde{U}^{(o)} h_{k} + b^{(o)} \right)\\
    \tilde { c }_{ t }  &= \tanh \left( W^{(c) } x _ { t } + U^{(c)}h_{t-1}  + \sum_{k \in P(t)} \tilde{U}^{(c)} h_{k} + b^{(c)} \right)\\
    f_{t,t-1} &= \sigma \left( W^{(f)} x_{t} + U^{(f)} h_{t-1} + b^{(f)} \right)\\
    f_{tk} &= \sigma \left( W^{(f)} x_{t} + \tilde{U}^{(f)} h_{k} + b^{(f)} \right), k\in P(t)\\
    c_{t}  &= i_{ t } \odot \tilde {c}_{t} +  f _ { t,t-1 } \odot c_{t-1} +  \sum _ { k \in P ( t ) } f _ { tk } \odot c_{k}\\
    h_{t}  &= o_{ t } \odot \tanh \left( c_{ t } \right)  
\end{align}


where $P(t)$ denotes the prior nodes that have \emph{real} edges linked to Node $t$.
Like the standard LSTM model (Equations 1-6), the proposed model also considers the hidden state of the immediate predecessor node ($t-1$) in the recurrent terms (i.e., $U^{(i)}$, $U^{(o)}$, $U^{(c)}$ and $U^{(f)}$), which correspond to the \emph{fake edges} discussed above.
\newtext{
However, when comparing the transition functions of both models (Equations \numrange{1}{6} and Equations~\numrange{7}{13}), it is easy to find the main difference: our model further considers the \emph{real edges} in the architecture and integrates the states of the remotely-connected predecessors (i.e., $\tilde{U} ^ {(i)}$, $\tilde{U} ^ {(o)}$, $\tilde{U} ^ {(c)}$ and $\tilde{U} ^ {(f)}$) into the current node. Thus, it can directly reflect the actual graph structure and well model the influence of former nodes on subsequent nodes along the node sequence in the graph drawing. 
}

\newtext{On the other hand, graphs are not sequential data.}
\newtext{When attempting to draw a graph using this approach, all nodes (both those before and those after in the linear layout) should be taken into consideration,}
i.e., \emph{the latter nodes in the sequence can also influence the positions of the former nodes during the actual graph drawing}. 
To better model this mutual influence, we further introduce a backward propagation to the proposed graph-LSTM-based model by simply reversing the link direction in the forward propagation (Fig.~\ref{fig_model}(b)). Then, we combine the outputs of each LSTM cell in both forward and backward propagations into a concatenated feature vector, which is further input into a fully-connected layer to generate the final 2D coordinate of each node.

\subsection{Model Input}
\label{sec_model_input}

When applying the LSTM model to graph drawing, it is crucial to find a suitable way to input the graph structure information into the LSTM model.
More specifically, we need to determine the \emph{feature vector} for each node, where the graph structure information of each node should be properly encoded. 
This feature vector will be further input into the LSTM model. Also, it is necessary to transform the original graph to a sequence of nodes (i.e., \emph{node ordering}) that can be processed by the LSTM model, which is another key point for applying LSTM into graph drawing.   

\textbf{Node Feature Vector:} When LSTM models are applied to NLP tasks, word embedding techniques are often used to transform words into fixed-length feature vectors; these vectors can be further input into LSTM models. Considering that many node embedding techniques have been proposed, like node2vec~\cite{grover2016node2vec}, DeepWalk~\cite{perozzi2014deepwalk} and SDNE~\cite{wang2016structural}, it is natural to use node embedding techniques to encode graph structure information and further input it into the proposed model. However, these node embedding techniques are mainly used for a single graph and have been proved incapable to be generalized to multiple graphs~\cite{heimann2017generalizing}. Our initial experiments during the design of the proposed method also confirmed this observation.

Taking into account that adjacency information between nodes is the essential information in a graph, we propose using a fixed-length adjacency vector as the feature vector of each node directly; it will be further input into the proposed model. The adjacency vector of each node encodes the connectivity between the current node and its prior $k$ nodes, where $k$ is empirically set as a fixed number.


\begin{figure}[b]
\centering 
\includegraphics[width=0.6\columnwidth]{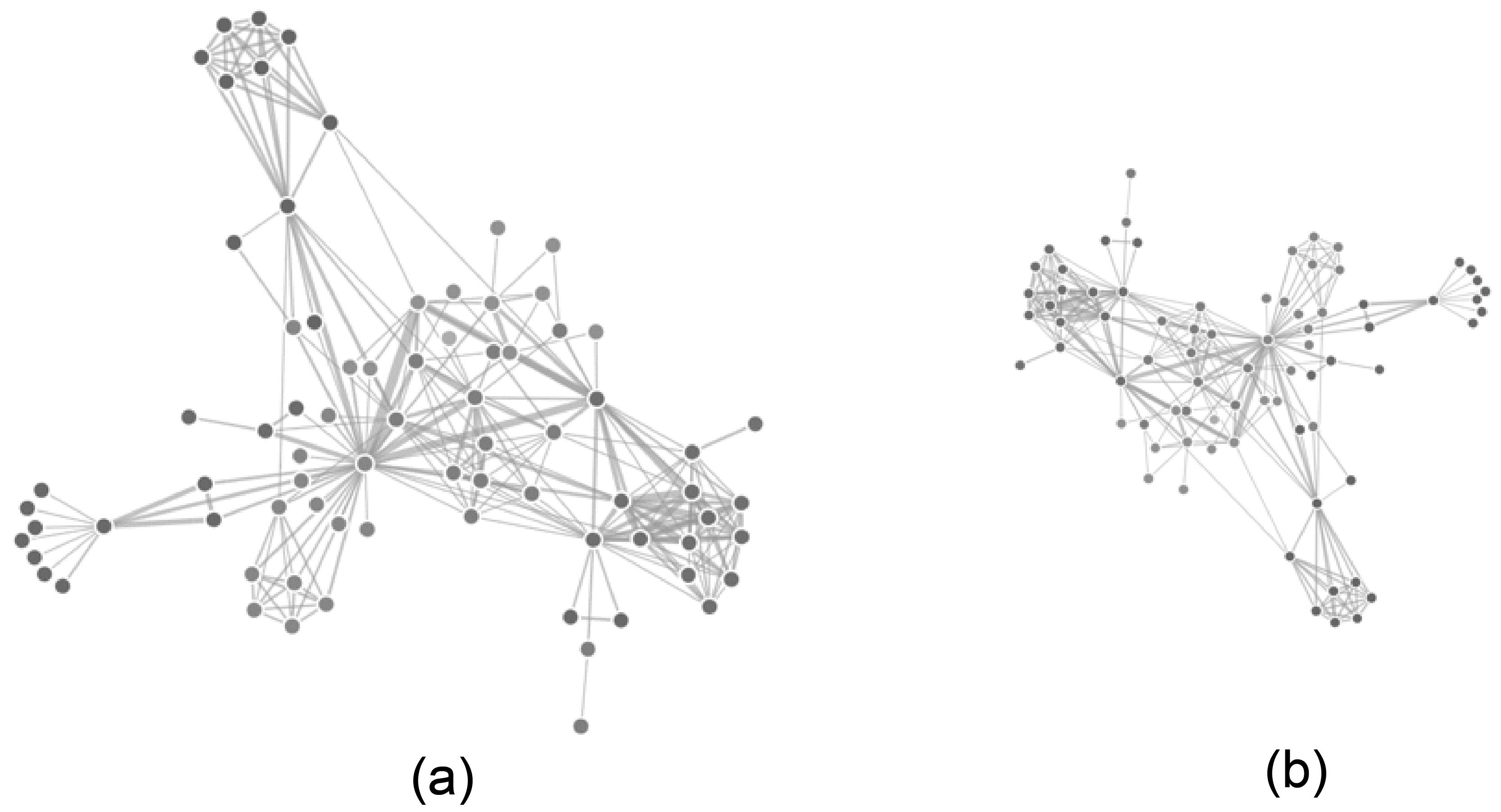}

\caption{The same graph drawing under transformations may look different: (a) the original graph drawing, (b) the same graph drawing that has been translated, rotated by 180 degrees and further scaled.}
\label{fig_drawing_transform}
\end{figure}

\textbf{Node Ordering:} Graphs can be represented by adjacency matrices and are permutation-invariant in terms of node ordering. Therefore, one naive way to model a graph as a sequence of nodes is to use random order. 
However, for a specific graph, the total number of such kind of random node orderings is $O(n!)$, where $n$ is the total number of nodes. 

\newtext{Inspired by the recent study~\cite{you2018graphrnn}, we propose using breadth-first-search (BFS) to generate node ordering for a specific graph. 
The major advantage of BFS ordering is that we need to train the proposed model on only all the possible BFS orderings, instead of exhaustively going through all possible node permutations,}
which can reduce the model searching space and benefit the model training.
In addition, for a specific node in a node sequence sorted by BFS, there is an upper bound for the possible connections between this node and the nodes before it along the BFS sequence~\cite{you2018graphrnn}.
More specifically, let $(v_1, v_2, ..., v_{i-1}, v_i, v_{i+1}, ...,v_n)$ be a BFS ordering of nodes, 
\newtext{
the furthest node before $v_i$ that is possible to link to $v_i$ is $v_{i-M}$ 
and all the nodes before $v_{i-M}$ are impossible to link to $v_i$,}
where $M$ is the maximum number of nodes of each level in a BFS sorting.
Due to the existence of this upper bound, we can set the length of the node feature vector to a fixed length smaller than the node number without losing much graph structure information.

To further reduce the model's searching space,
we also use node degree to sort the nodes at the same depth level of a BFS sorting.
During the training stage of the model, we randomly choose the starting node of the BFS sequence, which can augment the training dataset and improve model generalizability. 

\subsection{Loss Function}
\label{sec_loss}


The design of loss function is another crucial part of applying deep learning techniques to graph drawing, since the loss function will guide the neural network to learn 
\newtext{what
a desirable graph drawing
should be based on the training dataset.} Specifically,
the graph drawings in the training dataset are regarded as the ground-truth drawings, and the purpose of the loss function is to guide the proposed model to generate graph layouts as \textit{\textbf{``similar''}} to the corresponding training data as possible.

However, it is challenging to propose an appropriate loss function for comparing the similarity between two drawings of the same graph. 
\newtext{For example, a specific graph drawing may look very different after a series of operations from the set of translation, rotation and scaling,}
as shown in Fig.~\ref{fig_drawing_transform}. 
Therefore, the loss function should be invariant to such kind of transformations.
Motivated by these requirements, we propose conducting Procrustes Analysis~\cite{dryden2014shape,cox2000multidimensional} to assess the similarity between two drawings of the same graph. Procrustes Analysis is a widely-used technique in statistics and shape comparison~\cite{goodall1991procrustes} \newtext{and it has also been used in graph drawing~\cite{hu2012embedding}}. 
For two graph drawings of the same graph with $n$ nodes, Procrustes Analysis will explicitly translate, scale and rotate the drawings to align them.
Suppose the corresponding coordinates of all the $n$ nodes in the two drawings are $C=[c_1, ..., c_n]^T$ and $\bar{C} = [\bar{c}_1, ..., \bar{c}_n]^T$, where $c_i = (x_i, y_i)$ and $\bar{c}_i = (\bar{x}_i, \bar{y}_i)$, 
the \textbf{\textit{Procrustes Statistic}} will be calculated to indicate the shape difference between them as follows:
\begin{equation}
\label{eq_loss_func}
    R^2 = 1 - \frac{(tr(C^T\bar{C}\bar{C}^TC)^{1/2})^2}{tr(C^TC)tr(\bar{C}^T\bar{C})}
\end{equation}
where $0 \leq R^2 \leq 1$. It is essentially the squared sum of the distances between $C$ and $\bar{C}$ after a series of best possible transformations. 
$R^2 = 0$ denotes both graph drawings are exactly the same, while $R^2 = 1$ indicates that the two graph drawings are totally different and cannot be matched by any transformations.



\begin{table*}[h]
	\caption{Statistics of the generated graphs.}
	\centering
	\label{table_graph_statistics}
	\scalebox{0.9}{
	\begin{tabular}{c|c|c|c|c|c|c|c|c} \toprule
    Graph Type & \#Node & \#Edge & Node Degree     & Training & Validation & Testing &Total& \#Community     \\ \midrule
    Grid Graphs   & [100, 576] & [180, 1104] & [2,4] & 72  & 24 &24 & 120 & -   \\ \cmidrule{1-9}
    Star Graphs & [10, 209] & [9, 208] & [1,208] & 120  & 40 & 40 & 200 & -   \\ \midrule
    General Graphs   & [20, 50] & [23, 178] & [1,10] & 26000  & 3000 & 3000 & 32000 & [2, 12]   \\ \midrule
	\end{tabular}}
	\vspace{-4mm}
\end{table*}

\section{Evaluation}
\label{sec_evaluation}

We thoroughly assess {\name} through both qualitative and quantitative evaluations.
This section introduces the detailed experiment settings and evaluation results.

\subsection{Experiment Setup}





The experiment settings include graph generation, drawing dataset generation, baseline method selection and other implementation details.

\subsubsection{Graph Generation}
Since there are no public graph drawing datasets, where the layout position of each node is labeled,
we need to generate graphs and further properly draw the graphs to gain drawing datasets for training and evaluating {\name}.
For graph generation, we adopt the classical method of synthetic graph generation proposed by Lancichinetti et al.~\cite{lancichinetti2008graphGeneration},
\newtext{since it can generate realistic benchmark graphs with various community structures, which is helpful for verifying whether community-related visual properties are learned by {\name}.}
The implementation by the authors\footnote{\href{https://github.com/eXascaleInfolab/LFR-Benchmark\_UndirWeightOvp}{\url{https://github.com/eXascaleInfolab/LFR-Benchmark\_UndirWeightOvp}}} is used.
We can specify various parameters, including node number, average node degree, community number and community overlap coefficient, to control the structures of generated graphs.
In this paper, we mainly generated graphs with the node number evenly ranging from 20 to 50, as prior studies~\cite{huang2009nodeLinkEffect,ghoniem2004nodeLinkComparison} have shown that node-link diagrams are more suitable for graphs with dozens of nodes in terms of visual perception. 
We randomly split the whole dataset into training, validation and testing datasets. 
Table~\ref{table_graph_statistics} shows the detailed statistics of the generated graphs used in this paper.

Apart from general graphs, we also generate \newtext{grid and star graphs, which have simple and regular topological structures, to extensively evaluate the effectiveness of {\name}.}


To avoid contaminating the training data, we carefully checked and guaranteed that no validation and testing graphs have exactly the same topology structure with any training graphs by using pynauty
\footnote{\href{https://web.cs.dal.ca/~peter/software/pynauty/html}{\url{https://web.cs.dal.ca/~peter/software/pynauty/html}}},
a public library for checking graph isomorphism.




\subsubsection{Drawing Dataset Generation}
\label{subsec_drawingData}
We used different types of graph drawing methods to draw the graphs.
As discussed in Section~\ref{sec_graph_drawing}, there are mainly three types of graph drawing methods. 
\newtext{
There are many force-directed and dimension-reduction-based graph layout approaches.
We chose ForceAtlas2~\cite{jacomy2014forceatlas2}
and PivotMDS~\cite{brandes2006eigensolver} respectively, since both algorithms are typical and widely-used algorithms for each type.
Also, ForceAtlas2 can preserve the communities and PivotMDS is deterministic and fast.}
The multi-level drawing techniques are not used, as they mainly target at the acceleration of large graph visualization (Section~\ref{sec_graph_drawing}), which is not the focus of this paper.

All the graphs are drawn on a canvas with a size of $800 \times 800$, which is big enough for rendering the graphs used in this paper. 
When drawing the graphs using the above two methods, we manually tune the algorithm-specific parameters according to the node number, community structure and edge density of the input graphs, in order to achieve desirable visual properties \newtext{such as better clustering, fewer edge crossings and less node occlusion}. 
Also, due to the random point initialization of ForceAtlas2, its layout results are not deterministic.
Given a graph with the same algorithm-specific parameters, the graph drawing result may be different, which can confuse the deep learning model. 
Therefore, considering that PivotMDS is deterministic, we follow the method by Kruiger et al.~\cite{kruiger2017graph} and initialize the node positions with PivotMDS instead of using default randomized initial 2D positions, guaranteeing that 
the same input graph is mapped to the same layout.
By drawing the generated graphs using ForceAtlas2 and PivotMDS, we gained two drawing datasets with different layout styles.

For the grid and star graphs, apart from using ForceAtlas2 and PivotMDS, we also visualized grid graphs as perfect grid layout and star graphs as perfect star layout, gaining three drawing datasets \newtext{for grid and star graphs}.

\subsubsection{Baseline Method Selection}
To better evaluate the effectiveness of our approach, it is necessary to compare it with other general models that have a similar architecture. Initially, we compared unidirectional LSTM and bidirectional LSTM models, and also tested them with $1\sim4$ layers.
\newtext{Among all these general LSTM models, our initial results show that a 4-layer bidirectional LSTM (Bi-LSTM) model has the best graph drawing performance in terms of the Procrustes Statistic based similarity with the ground truth drawings.
This performance comparison result} is also consistent with Google Brain's prior study~\cite{britz2017lstm_testings}.
Therefore, a 4-layer bidirectional LSTM model is chosen as the baseline method for comparison in the subsequent evaluations.


\begin{table}[t]
	\caption{The configurations of the baseline model and our model.}
	\centering
	\label{table_model}
	\scalebox{0.9}{
	\begin{tabular}{c|c|c|c|c} \toprule
		       Model& Hidden Size & Layers  & Direction & \# Parameters    \\ \midrule
		       Baseline & 256 & 4 & bidirectional & 5.33M   \\ \midrule
              Ours & 256 & 1 & bidirectional & 1.12M   \\ \midrule
	\end{tabular}}
	\vspace{-4mm}
\end{table}

\subsubsection{Implementation and Model Configuration}
PivotMDS and ForceAtlas2 are implemented in Python based on
Tulip~\footnote{\href{http://tulip.labri.fr/Documentation/current/tulip-python/html/index.html}{\url{http://tulip.labri.fr/Documentation/current/tulip-python/html/index.html}}}
and
Gephi~\footnote{\href{https://github.com/bhargavchippada/forceatlas2}{\url{https://github.com/bhargavchippada/forceatlas2}}}. 
The proposed graph-LSTM-based model is implemented with PyTorch~\footnote{\href{https://pytorch.org/}{\url{https://pytorch.org/}}} and PyG Library~\footnote{\href{https://rusty1s.github.io/pytorch\_geometric}{\url{https://rusty1s.github.io/pytorch\_geometric}}}.
The LSTM model implementation integrated in PyTorch is used for the baseline model.
The machine we used for model training and all the subsequent experiments has 48 Intel Xeon(R) CPU processors (E5-2650 v4, 2.20GHz), and 4 NVIDIA Titan X (Pascal) GPUs.



\subsection{Model and Training Configurations}
The detailed configurations for both our model and the baseline model are shown in Table~\ref{table_model}.
\newtext{We use the Adam optimizer~\cite{kingma2014adam} for the model training.}
The learning rate and batch size for training both models are set to 0.0015 and 128, respectively.
The size of training, validation and testing graphs are shown in Table~\ref{table_graph_statistics}.

For each graph drawing dataset in Section~\ref{subsec_drawingData}, we train an individual model (the proposed graph-LSTM-based model or the baseline LSTM model) to learn the graph drawing style, with the corresponding algorithm-specific drawing parameters encoded in the model as well.
The size of the input adjacency vector of each node for both the grid and general graphs is empirically set as 35.
The star graphs are an extreme case, where the surrounding nodes are only connected to the center node.
Therefore, its input adjacency vector size is set as the maximum number of prior nodes, i.e., 208 (Table~\ref{table_graph_statistics}).
When the training loss converges, the corresponding model is used for generating graph drawings in the subsequent qualitative and quantitative evaluations.


\subsection{Qualitative Evaluation}
\label{sec_qe}


We first trained {\name} on the drawing datasets of \newtext{grid and star graphs}, where the graphs are drawn with three different drawing styles: perfect regular layouts (i.e., grid layout or star layout), \newtext{ForceAtlas2} and PivotMDS.
We further compared {\name} with the baseline model on the drawing datasets of general graphs with two drawing styles (i.e., ForceAtlas2 and PivotMDS).



	 
	 
	 
	 


\subsubsection{\newtext{Grid and Star Graphs}}

\begin{figure}[ht]
\centering 
\includegraphics[width=0.95\columnwidth]{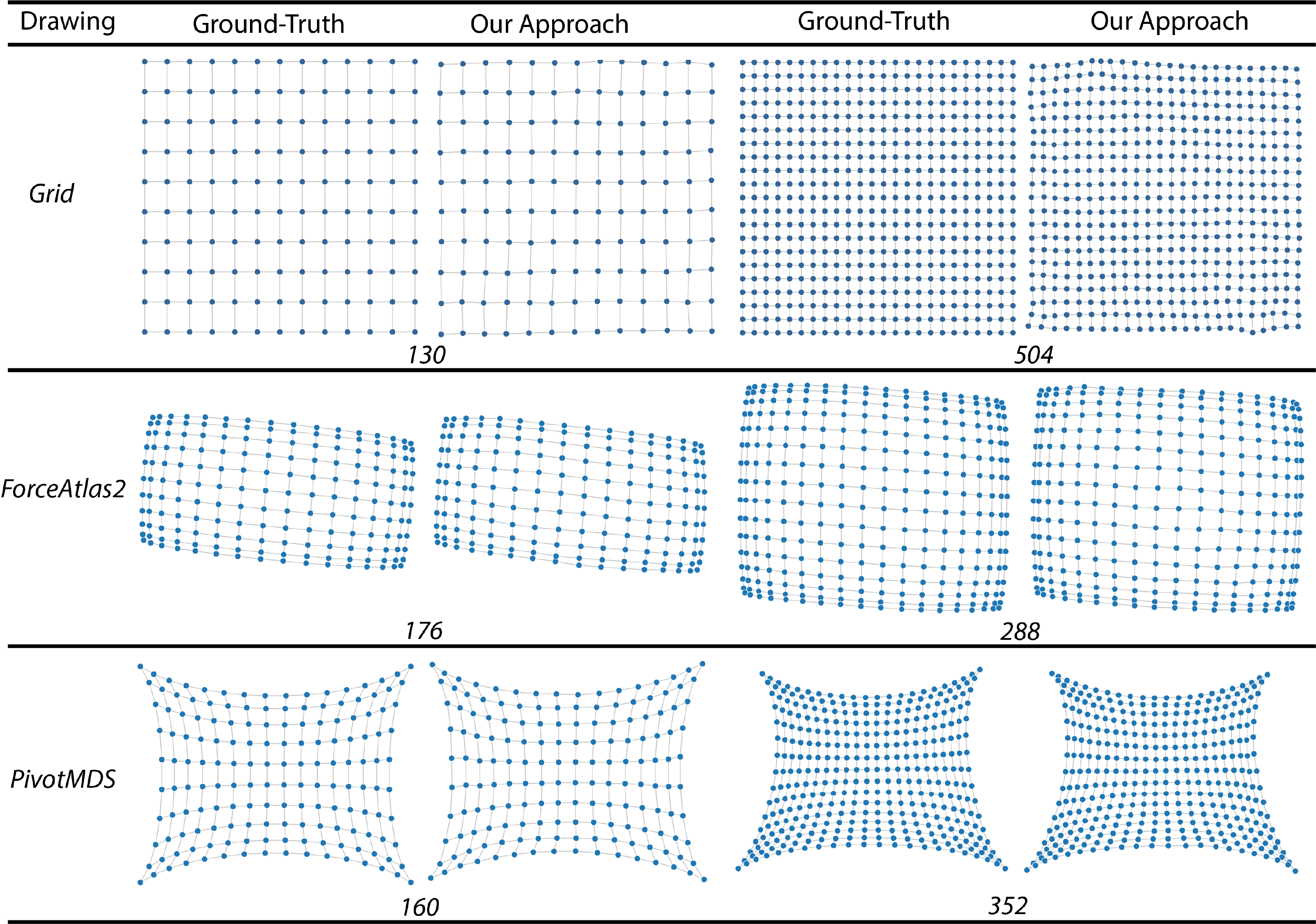}
\vspace{-2mm}
\caption{Qualitative evaluation results on grid graphs: the ground truth and the graph drawing generated by our approach are compared side by side. Each row shows the results of a specific drawing style (i.e., perfect grid layout, ForceAtlas2 and PivotMDS) and the number of graph nodes is shown in the bottom.}
\label{fig_case_grid}
\vspace{-2mm}
\end{figure}

Fig.~\ref{fig_case_grid} shows the graph drawing results of {\name} on grid graphs, where the drawings by {\name} and the corresponding ground truth drawings are aligned by the Procrustes Analysis.
The results demonstrate the excellent performance of {\name} in generating three different styles of graph drawings for grid graphs.
All the graph drawing styles in the ground truth layouts are well preserved by {\name}.
For example, the generated perfect grid layouts make all the nodes evenly distributed, but the results of {\name} trained on ForceAtlas2 drawings have a sparse distribution of nodes in the center and a dense distribution of nodes in the four corners, and results of {\name} trained on PivotMDS drawings tend to make the grid contours curved. All the generated graph drawings by our approach are consistent with the ground-truth drawings. 

\begin{figure}[h]
\centering 
\includegraphics[width=0.95\columnwidth]{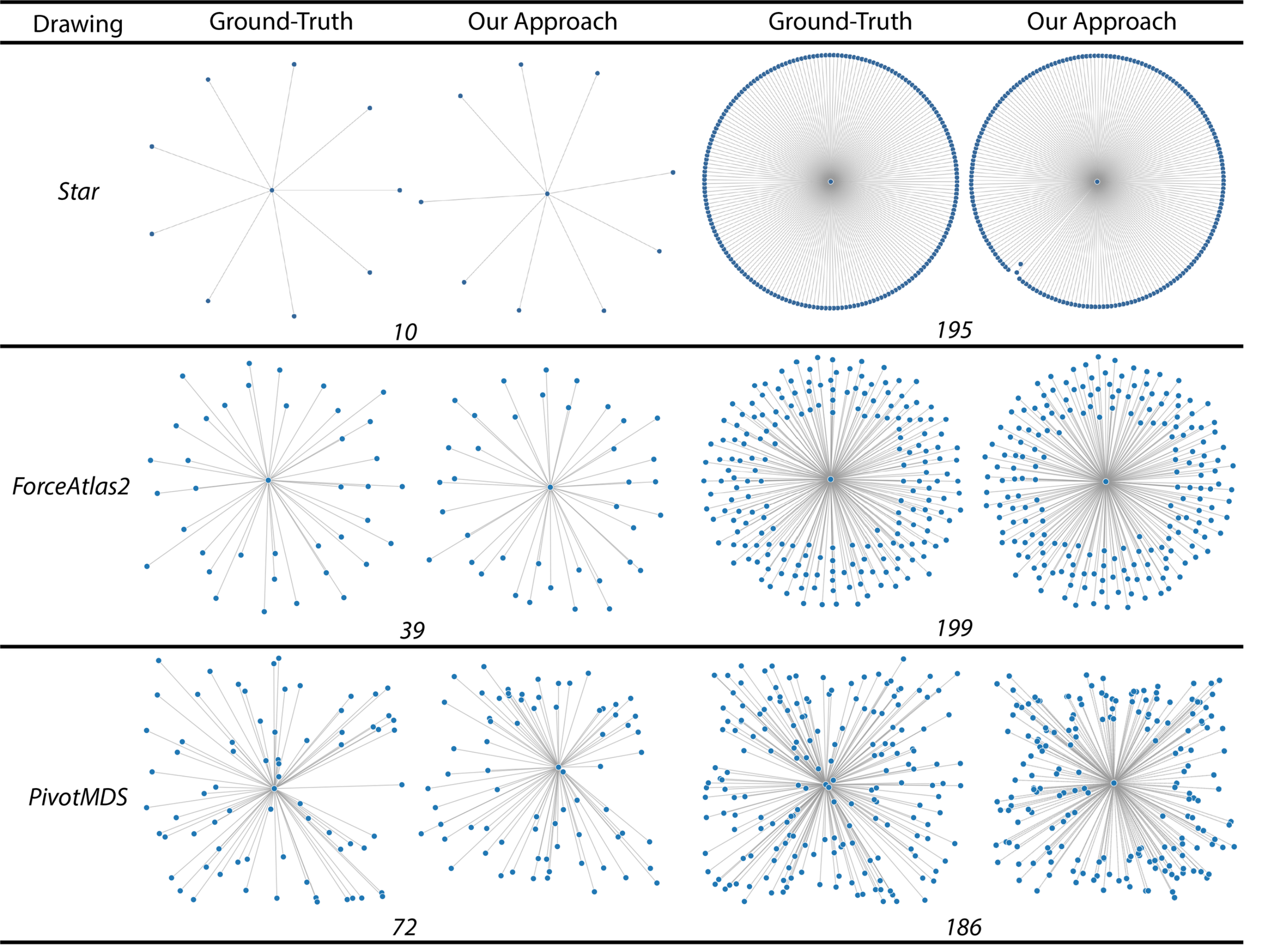}
\vspace{-2mm}
\caption{Qualitative evaluation results on star graphs: the ground truth and the graph drawing generated by our approach are compared side by side. Each row shows the results of a specific drawing style (i.e., perfect star layout, ForceAtlas2 and PivotMDS) and the number of graph nodes is shown in the bottom.}
\label{fig_case_star}
\vspace{-2mm}
\end{figure}

Fig.~\ref{fig_case_star} shows the drawing results of {\name} on the star graphs.
The drawing styles of perfect star layout, ForceAtlas2 and PivotMDS are different on star graphs, but {\name} is also able to learn these drawing styles and generates graph drawings that are very similar to the ground truth, further confirming the effectiveness of {\name}.

\subsubsection{General Graphs}

\begin{figure}[h]
\centering 
\includegraphics[width=0.95\linewidth]{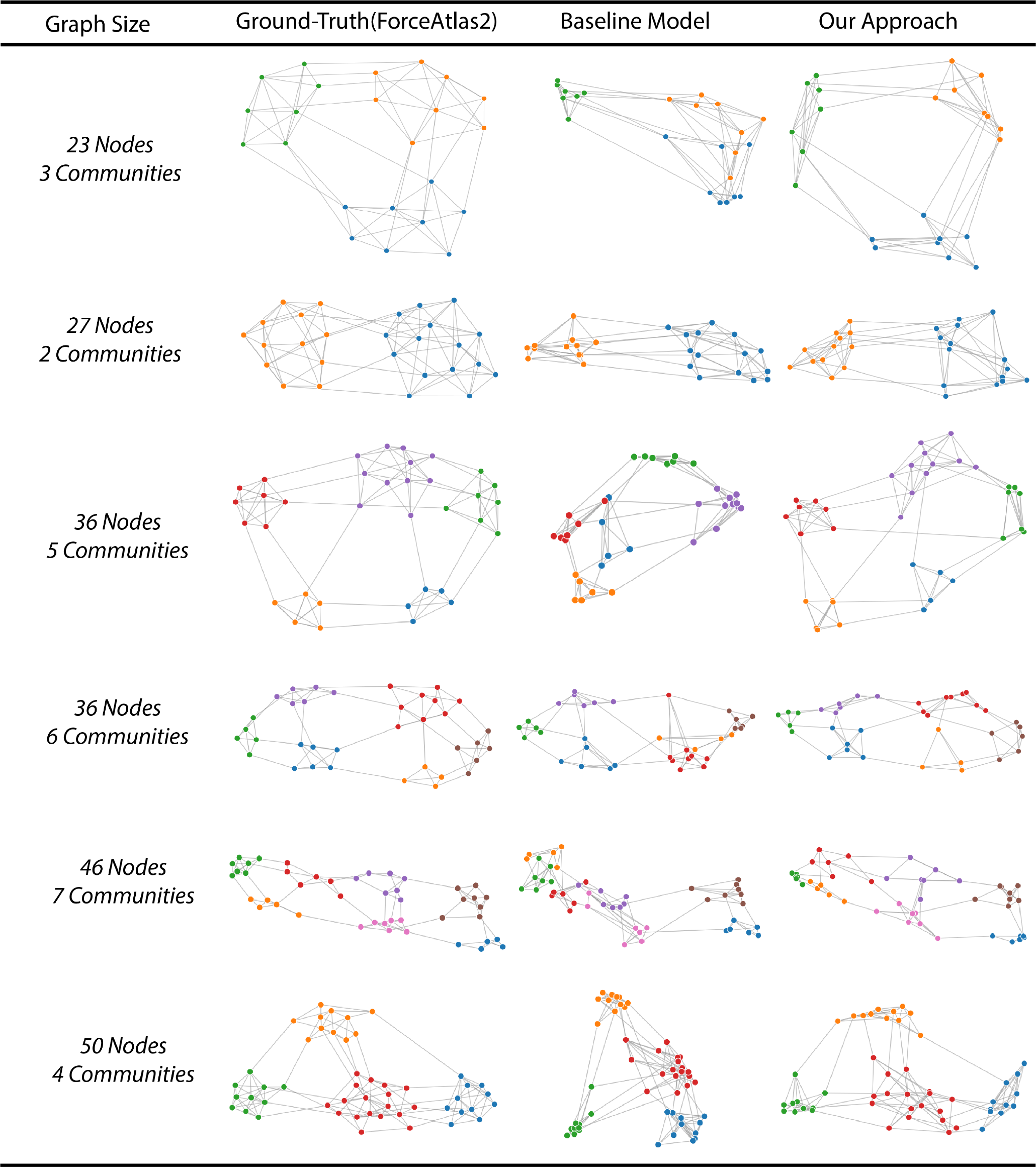}
\vspace{-2mm}

\caption{Qualitative evaluation on general graphs drawn by ForceAtlas2. For the same graph, the ground truth drawing, the drawing by the baseline model and the drawing by our approach are compared in each row.
Different colors indicate different communities.}
\label{fig_case_fa2}
\vspace{-4mm}
\end{figure}

Fig.~\ref{fig_case_fa2} shows the drawing results on the drawing dataset rendered by ForceAtlas2. The graph drawings generated by both the baseline method and our approach, as well as the ground-truth drawings by ForceAtlas2, are presented; these drawings cover different number of nodes and communities.
The ground-truth graph drawings (Column 2 of Fig.~\ref{fig_case_fa2}) can well visualize the community structures.
When comparing the drawing results of the baseline model (Column 3 of Fig.~\ref{fig_case_fa2}) with the ground-truth graph drawings, it is easy to see that the baseline model is able to preserve the overall community structures, especially when there are fewer nodes and communities (Rows 1 and 2 of Fig.~\ref{fig_case_fa2}). However, when there are more nodes and communities, the drawings generated by the baseline model deviate from the ground-truth drawings, and different communities may overlap with each other (Rows 3-6). 
On the contrary, our approach (Column 4 of Fig.~\ref{fig_case_fa2}) better preserves the community structures across different number of nodes and communities than the baseline method. 
When the drawing results by our approach are further compared to the corresponding ground-truth drawings, it is easy to see that our results reflect the visual properties of the ground-truth drawings and the overall node layouts in both drawings are similar.

\begin{figure}[h]
\centering 
\includegraphics[width=0.95\linewidth]{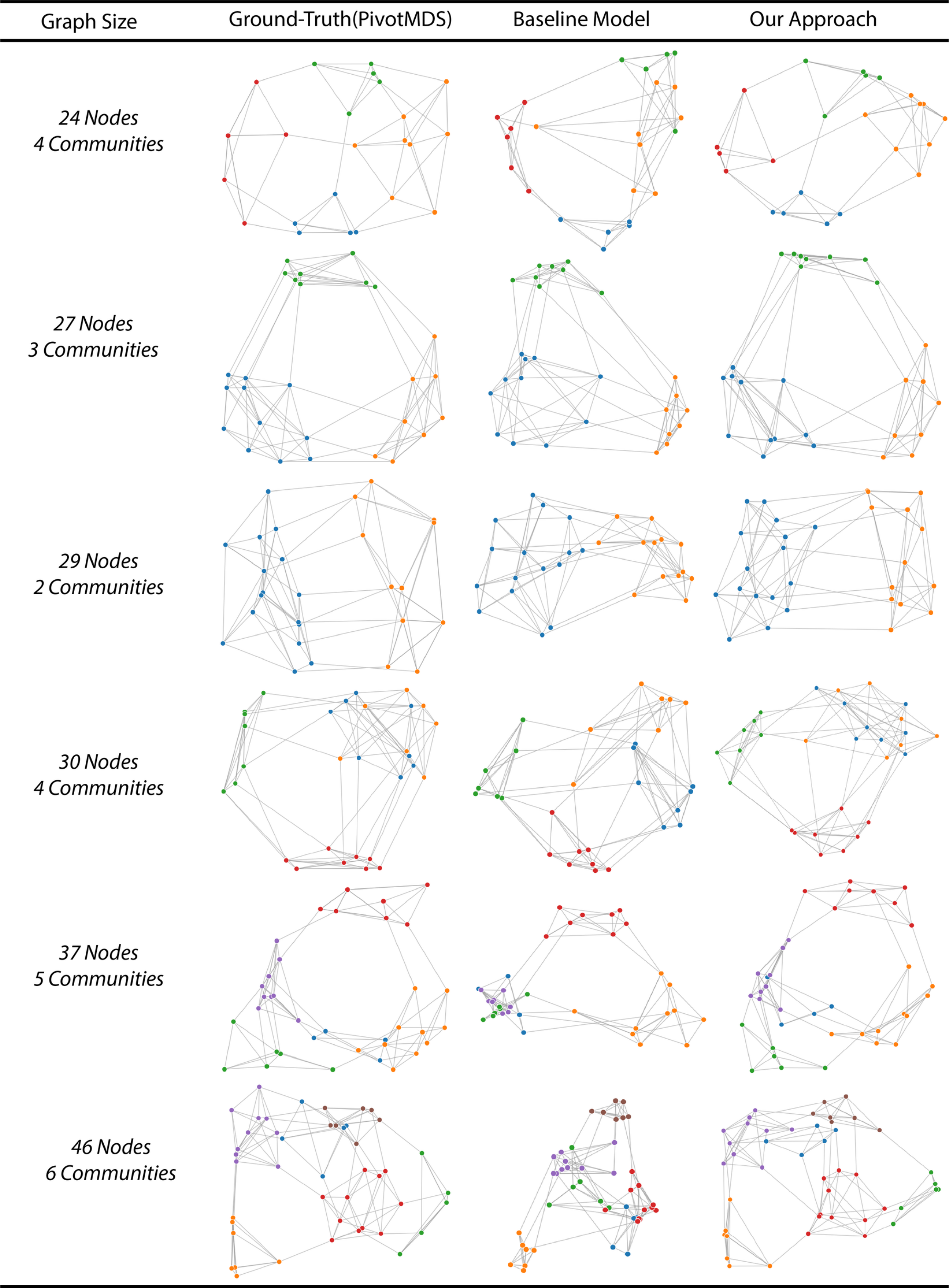}
\vspace{-2mm}
\caption{Qualitative evaluation on general graphs drawn by PivotMDS. For the same graph, the ground truth drawing, the drawing \newtext{generated} by the baseline model and the drawing \newtext{generated} by our approach are compared in each row.
Different colors indicate different communities.}
\label{fig_case_pivotMDS}
\vspace{-5mm}
\end{figure}

\newtext{
Fig.~\ref{fig_case_pivotMDS} shows the drawing results on the drawing dataset rendered by PivotMDS, where the ground-truth drawings by PivotMDS, the drawings generated by the baseline model and the drawings generated from our approach are compared side by side.}
Like the drawings of ForceAtlas2, the graph drawings of PivotMDS can also reflect the overall community structure of the graphs.
But there are also two major differences between these drawings in terms of the graph drawing style. One difference is that the edge length in the drawings of PivotMDS is more uniform than that of ForceAtlas2. 
Specifically, when compared with ForceAtlas2, the length of edges within communities is more similar to that of edges between different communities in PivotMDS layouts. 
The other difference is that some communities with tight connections can overlap each other in PivotMDS drawings, while the drawings of ForceAtlas2 often do not have community overlapping (Fig.~\ref{fig_case_fa2}).
For example, the light blue community and dark blue community overlap with each other in Row 4 of Fig.~\ref{fig_case_pivotMDS}, and the darker green community, light green community and darker blue community are also mixed in Row 6 of Fig.~\ref{fig_case_pivotMDS}, due to the tight connections between those communities.
\newtext{Such visual properties are well preserved by our approach.}
As shown in Column 4 of Fig.~\ref{fig_case_pivotMDS}, the community structure can be clearly recognized and the overlapping communities in the ground-truth drawings are also reflected (Rows 4 and 6). The overall layouts by our approach are quite similar to the ground-truth.
However, the baseline model (Column 3 of Fig.~\ref{fig_case_pivotMDS}) cannot preserve these visual properties (Rows 4-6), though it can reflect the community structures to some extent.


\subsection{Quantitative Evaluation}

We quantitatively evaluate the graph drawing results of both our approach and the baseline method by comparing their similarity with the ground-truth drawings.
This comparison is conducted from two perspectives: 
the Procrustes Statistic-based similarity and the aesthetic metrics-based similarity. 
The time costs of the graph drawing algorithms running on CPU and GPU are also reported.
\newtext{
This section considers only general graphs, 
since the qualitative evaluation has already shown that our approach achieves good performance on grid and star graphs.
}
All the subsequent experiments are conducted on the testing set of the general graph dataset (Table~\ref{table_graph_statistics}).

\subsubsection{Procrustes Statistic-based Similarity}
The Procrustes Statistic (Equation~\ref{eq_loss_func}), which was used as the loss function for the model training,
is further used to evaluate whether the models effectively learn a specific graph drawing style or not.
We analyzed the Procrustes Statistic-based similarity of all the testing graphs. We first ran Shapiro-Wilk test to check its normality, which indicates the results are not always normal. Thus, we further ran a Friedman test
with a Nemenyi-Damico-Wolfe-Dunn
for post-hoc analysis to determine the statistical significance (the statistical level $\alpha = 0.05$).

\begin{figure}[h]
\centering
\includegraphics[width=0.5\linewidth]{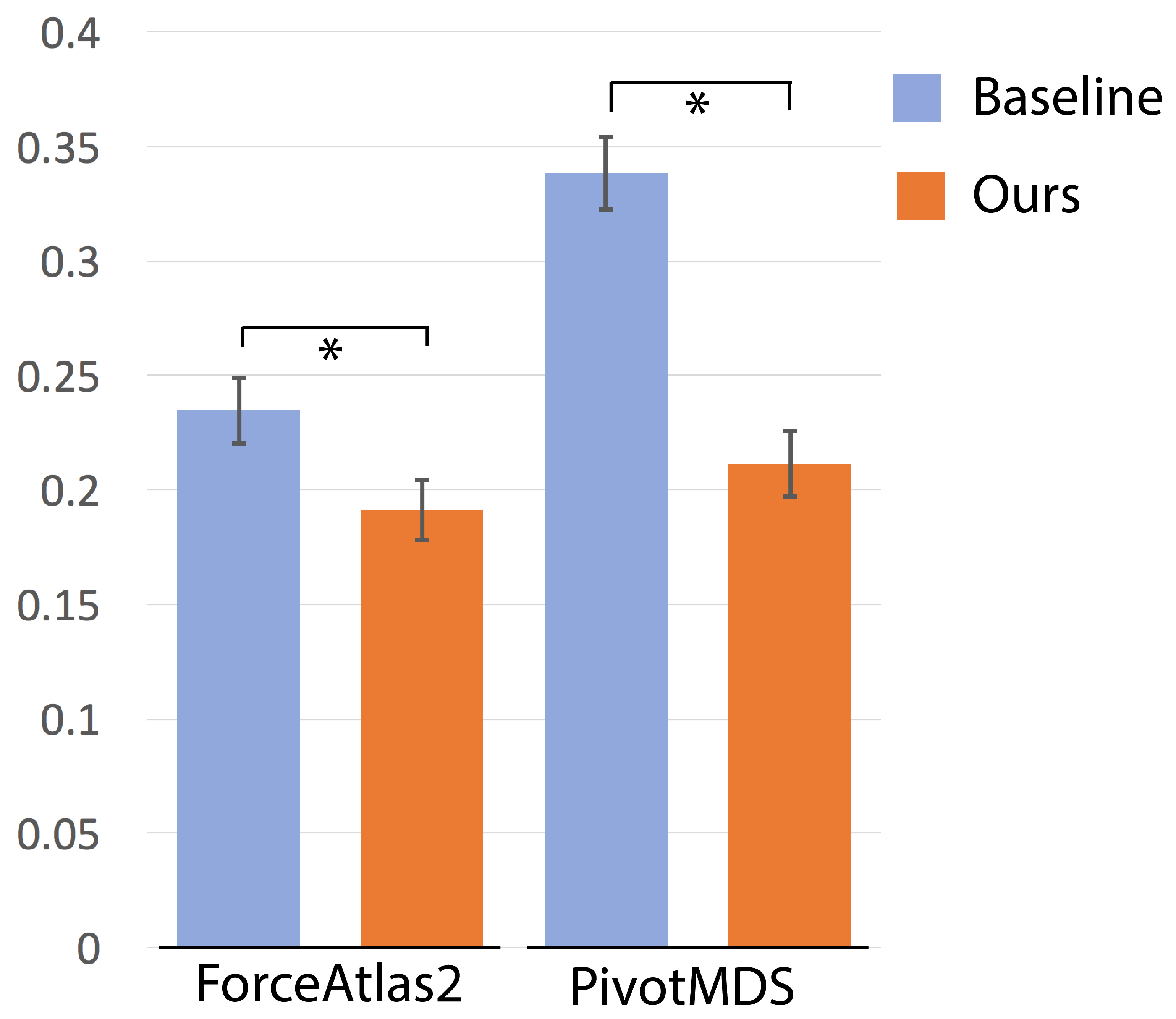}
\vspace{-2mm}
\caption{The results of Procrustes Statistic-based similarity. The baseline method and our approach are evaluated on both ForceAtlas2 and PivotMDS graph drawing datasets. The error bars are 95\% confidence
intervals and significant differences are marked with a line between them ($*$: $p < 0.05$). }
\label{fig_pa_difference}
\vspace{-2mm}
\end{figure}

Fig.~\ref{fig_pa_difference} shows the Procrustes Statistic-based similarity results, where the results on both ForceAtlas2 and PivotMDS drawing datasets are reported.
\newtext{
Compared with the baseline approach, our approach achieves a significantly better ($p < 0.05$) Procrustes Statistic-based similarity on both ForceAtlas2 (0.19 vs. 0.23) and PivotMDS (0.21 vs. 0.34) drawing dataset.}
This also provides support for the effectiveness of our approach in learning different graph drawing styles.

\subsubsection{Aesthetic Metrics-based Similarity}

\newtext{
Similar to the prior work~\cite{kwon2018would}, we also evaluate the effectiveness of our approach 
by comparing the aesthetic metric similarity between the drawings generated by our approach and the ground-truth drawings.
The \textbf{\textit{Root-Mean-Square Error (RMSE)}} is used to measure the aesthetic similarity between them. 
Given $n$ graphs, suppose the aesthetic metric values of the generated drawings are $\tilde{A} = \{\tilde{a}_1, ..., \tilde{a}_n \}$ and those of the ground-truth drawings are $A = \{a_1, ..., a_n \}$, then the similarity is defined as follows:}
\begin{equation}
    \operatorname{RMSE}(A, \tilde{A})=\sqrt{\frac{1}{n} \sum_{i}\left( \tilde{a}_{i} - a_{i} \right)^{2}}.
\end{equation}
\newtext{Smaller RMSE scores correspond to higher similarity to the ground-truth graph drawing.}

Various aesthetic metrics have been proposed to assess the aesthetic quality of different graph drawings~\cite{purchase2002metrics,dunne2015readability,wang2016ambiguityvis}.
\newtext{
In this paper, three aesthetic metrics are considered: edge crossings, node occlusions and community overlapping, since they are widely-used aesthetic criteria for evaluating how well the underlying graph topology has been realized in a drawing~\cite{dunne2009improving,purchase2002metrics,wang2016ambiguityvis}. Also, they have a normalized form and thus can be used to compare graphs of different sizes.
}

\emph{Edge crossings ($A_{ec}$):}
\newtext{We use the edge crossing metric introduced by Purchase~\cite{purchase2002metrics}, which is defined as the ratio of the number of edge crossings in a drawing over the upper bound of the possible crossings.}




\emph{Node occlusions ($A_{no}$):}
We choose the global metric of node occlusion introduced by Dune et al.~\cite{dunne2015readability}, i.e.,
\newtext{the ratio of the union area of all the node representations over their total area if drawn independently.}

\emph{Community overlapping ($A_{co}$):} We employ the global version of the autocorrelation-based metric introduced by Wang et al.~\cite{wang2016ambiguityvis}.
\newtext{For a specific node, this metric considers both the Euclidean distance between the node and its surrounding nodes and also whether they belong to the same community. Therefore, this metric can clearly reflect the degree of overlapping between different communities.}



Table~\ref{table_RSME_aesthetic} shows the results of the aesthetic metric-based similarities. For node occlusions, our approach is similar or slightly better than the baseline method on both the ForceAtlas2 and PivotMDS drawing datasets.
However, our approach has a better performance than the baseline method in terms of edge crossing similarity and community overlapping similarity. 
Since community structure can reflect the global graph structure, a better similarity of community structure indicates the better preservation of \newtext{a certain graph drawing style}.
Overall, the aesthetic metrics-based similarity comparisons confirm the effectiveness of our approach in preserving the original graph drawing \newtext{style}, which is also consistent with our observations in Section~\ref{sec_qe}.

\begin{table}[h]
	\caption{The \newtext{RMSEs} of aesthetic metrics-based similarity evaluated on the drawing datasets visualized by both ForceAtlas2 and PivotMDS.}
	\centering
	\label{table_RSME_aesthetic}
	\scalebox{0.95}{
	\begin{tabular}{c|c|c|c|c} \toprule
	 \multirow{2}{*}{Aesthetic Metrics} &  \multicolumn{2}{c|}{ForceAtlas2} & \multicolumn{2}{c}{PivotMDS}  \\ \cmidrule{2-5}
	 
	 & Baseline & Ours & Baseline & Ours \\ \midrule
	 $RMSE(A_{ec})$ & 0.0169 & \textbf{0.0125} & 0.0191 & \textbf{0.0134} \\ \midrule
	 $RMSE(A_{no})$ & 0.0316 & 0.0310 & 0.0799 & 0.0794 \\ \midrule
	 $RMSE(A_{co})$ & 0.0138 & \textbf{0.0125} & 0.0171 & \textbf{0.0131} \\ \midrule
	\end{tabular}}
	\vspace{-5mm}
\end{table}

\subsubsection{Time Cost}
We evaluate the time cost of the proposed approach in comparison with both the original graph drawing techniques (i.e., ForceAtlas2 and PivotMDS) and the baseline model. 
The time costs on both CPU and GPU are tested.
For CPU mode, each graph is repeatedly drawn 10 times. 
Their average is regarded as the actual time cost of drawing a graph.
The iteration number for ForceAtlas2 is empirically set as 700 steps, since the graph drawing results after 700 steps are relatively stable for the given graphs, though it is indicated that it can achieve a slightly better drawing quality with more steps~\cite{jacomy2014forceatlas2}.
For GPU mode, we only compared the time cost of our approach with that of the baseline model to provide a quick understanding of how fast the deep learning-based approaches can achieve on GPUs. 
Since the major advantages of GPU-based programming lie in its parallel computation, the testing graphs are input into the model all together to generate the drawings and the corresponding average time cost for each graph is calculated accordingly.

\begin{figure}[!t] 
\centering 
\includegraphics[width=0.5\linewidth]{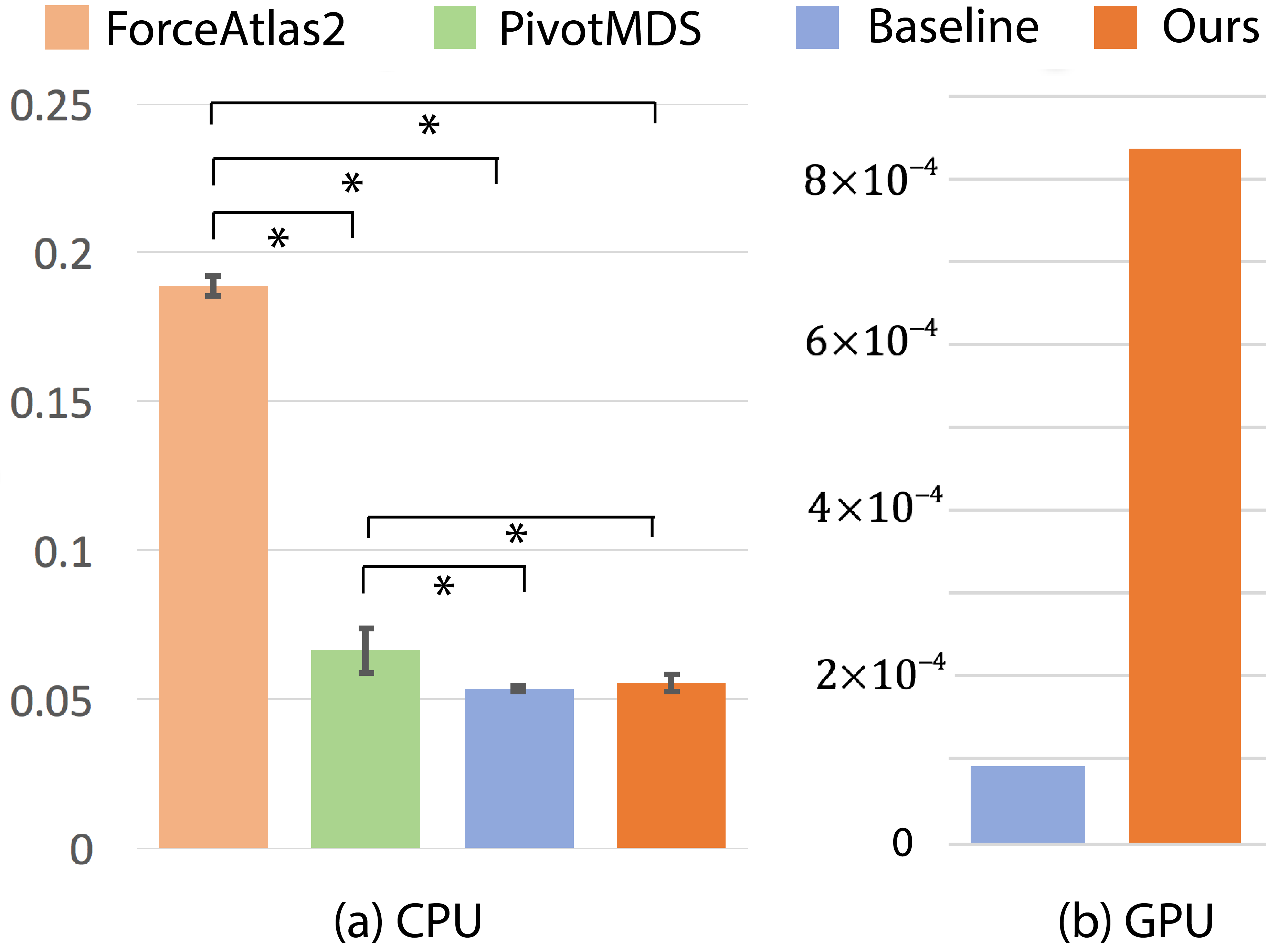}
\vspace{-2mm}
\caption{A comparison of the average running time (second) for drawing each graph using different methods. (a) The average running time on CPU, where the error bars are 95\% confidence
intervals and the graph drawing techniques with significant difference are marked with a line between them ($*$: $p < 0.05$), (b) the average running time of the baseline method and our approach on GPU.}
\label{fig_case_timeCost}
\vspace{-4mm}
\end{figure}

Fig.~\ref{fig_case_timeCost} shows the average time cost of using the graph drawing methods to draw a graph on both CPU and GPU.
The CPU time costs do not follow a normal distribution according to a Shapiro-Wilk test. Thus, we ran a
non-parametric Friedman test with a Nemenyi-Damico-Wolfe-Dunn
for post-hoc analysis to determine the statistical significance ($\alpha = 0.05$). 
Fig.~\ref{fig_case_timeCost}(a) shows that our approach and the baseline approach have a similar CPU time cost (0.055 vs. 0.054 second) and both of them are significantly faster than the original graph drawing methods on CPU. Specifically, the time cost of our approach is only 29\% of ForceAtlas2 (0.189 second) and 82\% of PivotMDS (0.067 second). 
Fig.~\ref{fig_case_timeCost}(b) indicates that our approach ($8.3 \times 10^{-4}$ second) is slower than the baseline method ($0.93 \times 10^{-4}$ second) on GPU,
\newtext{though its training parameter number is only about $20\%$ of the baseline model (Table~\ref{table_model}).
One major reason for this is that we directly used the implementation integrated in PyTorch for the baseline LSTM model, whose underlying implementation is based on C/C++ and has been extensively optimized with NVIDIA cuDNN, but the implementation of {\name} is Python without optimization using NVIDIA cuDNN.}
However, the speed of both deep learning-based methods is increased by two orders of magnitude when running on GPU than that of the techniques running on CPU.

\begin{figure}[!t] 
\centering 
\includegraphics[width=0.6\linewidth]{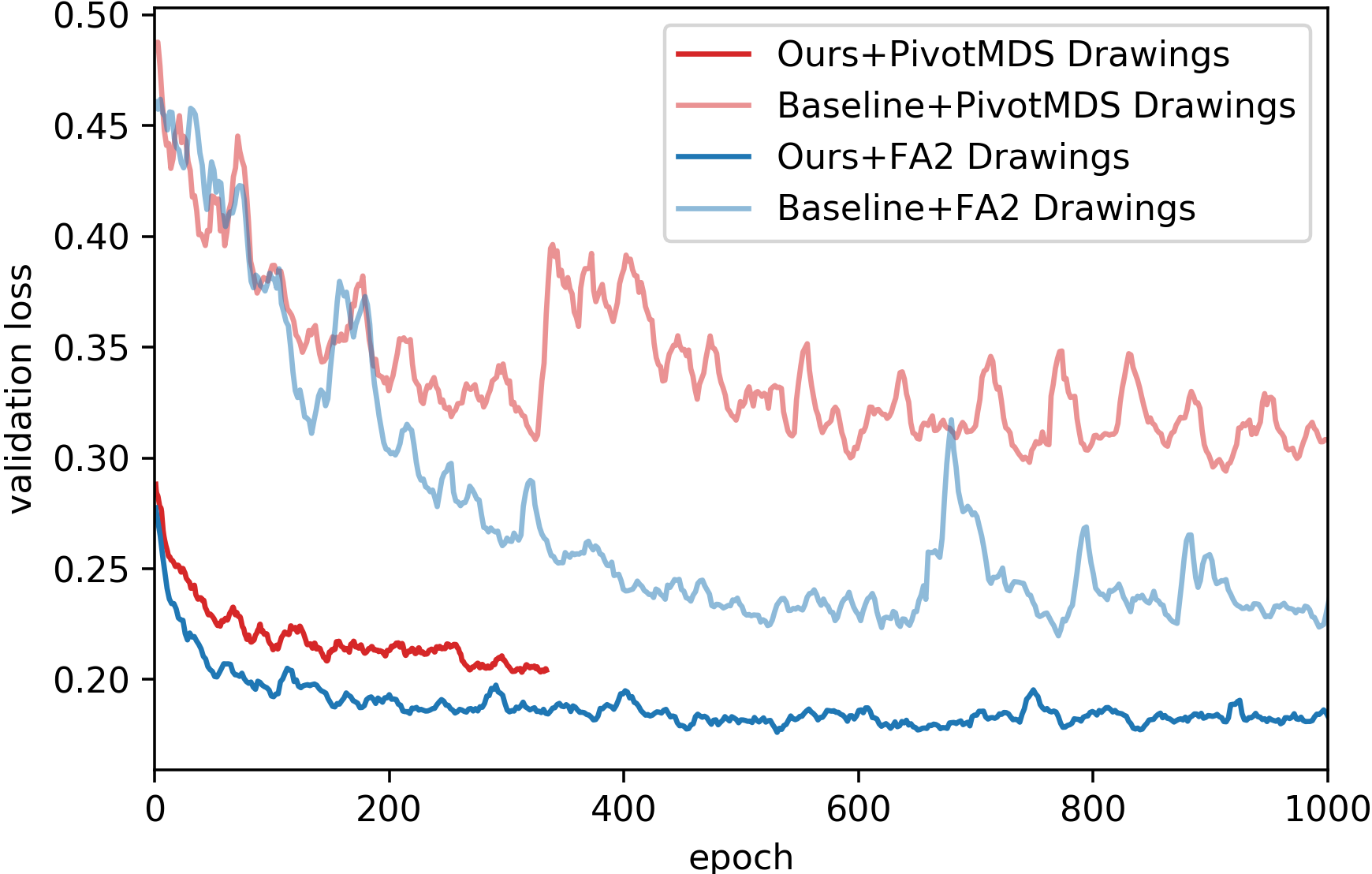}
\vspace{-2mm}
\caption{The loss curves of training the baseline and our model on ForceAtalas2 and PivotMDS drawing datasets. The training of our model on PivotMDS drawings is stopped at 330 epochs due to its convergence.}
\label{fig_training_loss}
\vspace{-4mm}
\end{figure}

\newtext{For the model training, Fig.~\ref{fig_training_loss} shows the training validation loss curves.
It is easy to observe that {\name} needs fewer epochs to converge and has lower and smoother training loss than the baseline model.
One possible explanation for this is that the architecture of {\name} explicitly considers the input graph structure.
}


\section{Discussions}
\newtext{In this section, we first report the lessons we learned when working on this paper. Then, we further discuss the limitations and usage issues of the proposed approach for graph drawing.}

\subsection{\newtext{Lessons}}
\newtext{We learned many lessons from the model design and training while working on this paper.}

\textbf{Node Feature Vector}
For determining the adjacency vector size, a naive choice is to set it as the maximum node number of all the graphs, which, however, is at the cost of model complexity and efficiency.
We quickly evaluated its influence and found that there is no significant degradation of drawing performance for general graphs, with the decreasing of node size in a certain range. 
This is probably because the architecture of {\name} explicitly encodes the graph structure, and there is an upper bound of the distance between connected nodes in a BFS sequence of a graph~\cite{you2018graphrnn}.

\textbf{Model Hyper-parameters}
The choice of model hyper-parameters (e.g., hidden size, layer numbers) of {\name} is crucial for deep learning models. With the increase of these hyper-parameters, the learning capability of the model can often increase, but it is also at the cost of increasing overfitting risk and training time and decreasing model efficiency.
For {\name}, we conducted control experiments to
\newtext{balance}
these factors to determine the suitable hyper-parameters.
For example, we found that increasing the layer number to 2 or 3 brings little improvement in the drawing performance, but changing the model to be unidirectional results in significant performance degeneration.

\textbf{Loss Function Design}
For the loss function, we also \newtext{considered} other possible choices, for example, the sum of the edge length difference between the predicted drawing and the ground truth.
This kind of loss function can delineate the similarity of different drawings under rotation and translation. However, it cannot capture the similarity between two drawings under 
scaling
and the loss value seems to be dominated by long edges.
After a series of careful designs and comparisons, the Procrustes Statistic was finally chosen to guide the model training.

\textbf{Drawing Complexity, Training Size and Overfitting}
As shown in Table~\ref{table_graph_statistics}, we used a small training dataset for grid and star graphs (72 for grid graphs and 120 for star graphs), the graph drawing performance is good and no overfitting was observed. The major reason for this is that the graph drawings for grid and star graphs have fixed patterns with fewer variations, which makes it easier for the deep learning model to learn from them.
On the contrary, the drawings of general graphs are complex. Initially, we used about 8000 graphs with a fixed order of input nodes for training on general graphs, which results in an overfitted model. 
This is solved by randomly selecting the starting node of the input BFS-ordering node sequence, which essentially augmented the training data, and further increased the training data size.


\subsection{\newtext{Limitations and Usage Issues}}
\newtext{Our evaluations above have shown that {\name} can learn from the training graph drawings (i.e., grid layout, star layout, ForceAtlas2 and PivotMDS) and further generate drawings
for new graph inputs.
However, the proposed method has limitations and next we discuss some of them.
}

\newtext{
\textbf{Failure Cases and Limited Evaluations} According to our empirical observations, two factors can affect the performance of {\name} in generating drawings with a drawing style similar to the training drawing: \emph{graph structure similarity} and \emph{node ordering}. When the input graph has significantly-different graph structures or the nodes are not sorted by BFS, it can result in a decrease of drawing performance of {\name} or even generate messy drawings.
Also, as the first step of using deep learning for graph drawing, {\name} currently focuses on small graphs (i.e., clustered graphs with \numrange{20}{50} nodes).
Testing whether the proposed approach generalizes to large graphs requires further exploration.
}


\newtext{
\textbf{Model Interpretability and Interactivity} {\name} is a deep learning based approach. 
Like most of deep learning based methods, it also has the interpretability issues, which are an active research topic in both visualization and machine learning~\cite{zhang2018visual,hohman2018visual,murdoch2019interpretable}. Specifically, in the case of {\name}, it is not clear what graph layout aspects are learned by the neural network.
Also, once the model training is done, {\name} can directly generate drawings with a drawing style similar to the training data for input graphs, where no user interaction is needed. This is both an advantage and disadvantage. It can benefit novice users without a background in graph drawing, but may be a disadvantage for expert users who want to interactively explore more graph properties by themselves.}

\newtext{
\textbf{Usage Issues}
With a trained model for a specific type of graph and a specific drawing style, {\name} can be used to generate similar drawings of similar graphs. However, many sample graphs and the drawings of these graphs are needed, as well as expert interaction with the model (e.g., hyper-parameter tuning), and the time to train the model. In order to generate different types of drawings, the model needs to be retrained on new graphs and new graph drawings.
}





\section{Conclusion}

In this paper, we propose {\name}, a novel graph-LSTM-based approach for graph drawing, where the graph drawing is formalized as a learning and prediction problem. 
\newtext{
Given a graph drawing dataset, {\name} is trained to learn a graph drawing style and can further generate graph drawings with similar characteristics.
We carefully designed the proposed approach in terms of model architecture, model input and training loss.
We conducted both qualitative and quantitative evaluations on three types of graphs (i.e., grid graphs, star graphs and general graphs with good community structures) and four types of drawings (i.e., grid layout, star layout, ForceAtlas2 and PivotMDS).
The results show that {\name} can generate similar drawings for the three types of graphs and its speed is fast on the testing graphs with \numrange{20}{50} nodes, which provides support for the effectiveness and efficiency of {\name} for graph drawing. Also, it is observed that {\name} 
can better preserve the original graph drawing style than the general LSTM-based method, confirming the advantage of our model architecture.
}


In future work, we plan to optimize the current implementation of {\name} (e.g., accelerations with PyTorch C/C++ extensions and NVIDIA cuDNN) and further evaluate its performance on additional types of graphs. 
\newtext{Also, it would be interesting to explore how a deep learning approach can benefit the visualization of large graphs with thousands of nodes. For example, given our comparison results of time cost on small graphs with 20 to 50 nodes, deep learning based approaches may also be able to improve the efficiency of large graph drawing.}
Furthermore, since dynamic graph visualization often depends on the temporal correlation between adjacent time stamps,
it is also promising to investigate whether deep learning techniques can be extended to dynamic graph visualization.
\newtext{We hope this work can inspire more research on using deep learning techniques for graph drawing as well as general information visualization.}









    
      
      

\acknowledgments{
The authors wish to thank Yao Ming, Qiaomu Shen and Daniel Archambault for the constructive discussions. The authors also thank the anonymous reviewers for their valuable comments.
This work is partially supported by a grant from MSRA (code: MRA19EG02).
}

\bibliographystyle{abbrv-doi}
\balance
\bibliography{reference}

\begin{thebibliography}{10}

\bibitem{atwood2016diffusion}
J.~Atwood and D.~Towsley.
\newblock Diffusion-convolutional neural networks.
\newblock In {\em Proceedings of Advances in Neural Information Processing
  Systems}, pp. 1993--2001, 2016.

\bibitem{bach2012interactive}
B.~Bach, A.~Spritzer, E.~Lutton, and J.-D. Fekete.
\newblock Interactive random graph generation with evolutionary algorithms.
\newblock In {\em International Symposium on Graph Drawing}, pp. 541--552.
  Springer, 2012.

\bibitem{barbosa2001interactive}
H.~J. Barbosa and A.~Barreto.
\newblock An interactive genetic algorithm with co-evolution of weights for
  multiobjective problems.
\newblock In {\em Proceedings of the 3rd Annual Conference on Genetic and
  Evolutionary Computation}, pp. 203--210. Morgan Kaufmann Publishers Inc.,
  2001.

\bibitem{battista1998graph}
G.~D. Battista, P.~Eades, R.~Tamassia, and I.~G. Tollis.
\newblock {\em Graph drawing: algorithms for the visualization of graphs}.
\newblock Prentice Hall PTR, 1998.

\bibitem{bengio1994learning}
Y.~Bengio, P.~Simard, P.~Frasconi, et~al.
\newblock Learning long-term dependencies with gradient descent is difficult.
\newblock {\em IEEE Transactions on Neural Networks}, 5(2):157--166, 1994.

\bibitem{biedl1998graph}
T.~Biedl, J.~Marks, K.~Ryall, and S.~Whitesides.
\newblock Graph multidrawing: Finding nice drawings without defining nice.
\newblock In {\em International Symposium on Graph Drawing}, pp. 347--355.
  Springer, 1998.

\bibitem{bonabeau2002graph}
E.~Bonabeau.
\newblock Graph multidimensional scaling with self-organizing maps.
\newblock {\em Information Sciences}, 143(1-4):159--180, 2002.

\bibitem{bonabeau1998self}
E.~Bonabeau and F.~H{\'e}naux.
\newblock Self-organizing maps for drawing large graphs.
\newblock {\em Information Processing Letters}, 67(4):177--184, 1998.

\bibitem{brandes2006eigensolver}
U.~Brandes and C.~Pich.
\newblock Eigensolver methods for progressive multidimensional scaling of large
  data.
\newblock In {\em International Symposium on Graph Drawing}, pp. 42--53.
  Springer, 2006.

\bibitem{bresson2018experimental}
X.~Bresson and T.~Laurent.
\newblock An experimental study of neural networks for variable graphs.
\newblock 2018.

\bibitem{britz2017lstm_testings}
D.~Britz, A.~Goldie, M.-T. Luong, and Q.~Le.
\newblock Massive exploration of neural machine translation architectures.
\newblock {\em arXiv preprint arXiv:1703.03906}, 2017.

\bibitem{bruna2014spectral}
J.~Bruna, W.~Zaremba, A.~Szlam, and Y.~Lecun.
\newblock Spectral networks and locally connected networks on graphs.
\newblock In {\em Proceedings of International Conference on Learning
  Representations}, 2014.

\bibitem{cimikowski1996neural}
A.~Cimikowski and P.~Shope.
\newblock A neural-network algorithm for a graph layout problem.
\newblock {\em IEEE Transactions on Neural Networks}, 7(2):341--345, 1996.

\bibitem{cox2000multidimensional}
T.~F. Cox and M.~A. Cox.
\newblock {\em Multidimensional scaling}.
\newblock Chapman and hall/CRC, 2000.

\bibitem{defferrard2016gcn}
M.~Defferrard, X.~Bresson, and P.~Vandergheynst.
\newblock Convolutional neural networks on graphs with fast localized spectral
  filtering.
\newblock In {\em Proceedings of Advances in Neural Information Processing
  Systems}, pp. 3844--3852, 2016.

\bibitem{di1994algorithms}
G.~Di~Battista, P.~Eades, R.~Tamassia, and I.~G. Tollis.
\newblock Algorithms for drawing graphs: an annotated bibliography.
\newblock {\em Computational Geometry-Theory and Application}, 4(5):235--282,
  1994.

\bibitem{dos2015application}
R.~dos Santos~Vieira, H.~A.~D. do~Nascimento, and W.~B. da~Silva.
\newblock The application of machine learning to problems in graph drawing a
  literature review.
\newblock In {\em Proceedings of International Conference on Information,
  Process, and Knowledge Management}, pp. 112--118, 2015.

\bibitem{dryden2014shape}
I.~L. Dryden.
\newblock Shape analysis.
\newblock {\em Wiley StatsRef: Statistics Reference Online}, 2014.

\bibitem{dunne2015readability}
C.~Dunne, S.~I. Ross, B.~Shneiderman, and M.~Martino.
\newblock Readability metric feedback for aiding node-link visualization
  designers.
\newblock {\em IBM Journal of Research and Development}, 59(2/3):14--1, 2015.

\bibitem{dunne2009improving}
C.~Dunne and B.~Shneiderman.
\newblock Improving graph drawing readability by incorporating readability
  metrics: A software tool for network analysts.
\newblock {\em University of Maryland, HCIL Tech Report HCIL-2009-13}, 2009.

\bibitem{duvenaud2015convolutional}
D.~K. Duvenaud, D.~Maclaurin, J.~Iparraguirre, R.~Bombarell, T.~Hirzel,
  A.~Aspuru-Guzik, and R.~P. Adams.
\newblock Convolutional networks on graphs for learning molecular fingerprints.
\newblock In {\em Proceedings of Advances in Neural Information Processing
  Systems}, pp. 2224--2232, 2015.

\bibitem{eades1984heuristic}
P.~Eades.
\newblock A heuristic for graph drawing.
\newblock {\em Congressus Numerantium}, 42:149--160, 1984.

\bibitem{espadoto2019deep}
M.~Espadoto, N.~S. Hirata, and A.~C. Telea.
\newblock Deep learning multidimensional projections.
\newblock {\em arXiv preprint arXiv:1902.07958}, 2019.

\bibitem{frick1994fast}
A.~Frick, A.~Ludwig, and H.~Mehldau.
\newblock A fast adaptive layout algorithm for undirected graphs (extended
  abstract and system demonstration).
\newblock In {\em International Symposium on Graph Drawing}, pp. 388--403.
  Springer, 1994.

\bibitem{fruchterman1991graph}
T.~M. Fruchterman and E.~M. Reingold.
\newblock Graph drawing by force-directed placement.
\newblock {\em Software: Practice and experience}, 21(11):1129--1164, 1991.

\bibitem{gajer2000multi}
P.~Gajer, M.~T. Goodrich, and S.~G. Kobourov.
\newblock A multi-dimensional approach to force-directed layouts of large
  graphs.
\newblock In {\em International Symposium on Graph Drawing}, pp. 211--221.
  Springer, 2000.

\bibitem{gansner2004graph}
E.~R. Gansner, Y.~Koren, and S.~North.
\newblock Graph drawing by stress majorization.
\newblock In {\em International Symposium on Graph Drawing}, pp. 239--250.
  Springer, 2004.

\bibitem{gers1999learning}
F.~A. Gers, J.~Schmidhuber, and F.~A. Cummins.
\newblock Learning to forget: Continual prediction with lstm.
\newblock {\em Neural Computation}, 12:2451--2471, 2000.

\bibitem{ghoniem2004nodeLinkComparison}
M.~Ghoniem, J.-D. Fekete, and P.~Castagliola.
\newblock A comparison of the readability of graphs using node-link and
  matrix-based representations.
\newblock In {\em IEEE Symposium on Information Visualization}, pp. 17--24,
  2004.

\bibitem{gibson2013survey}
H.~Gibson, J.~Faith, and P.~Vickers.
\newblock A survey of two-dimensional graph layout techniques for information
  visualisation.
\newblock {\em Information Visualization}, 12(3-4):324--357, 2013.

\bibitem{goodall1991procrustes}
C.~Goodall.
\newblock Procrustes methods in the statistical analysis of shape.
\newblock {\em Journal of the Royal Statistical Society: Series B
  (Methodological)}, 53(2):285--321, 1991.

\bibitem{goodfellow2016deep}
I.~Goodfellow, Y.~Bengio, A.~Courville, and Y.~Bengio.
\newblock {\em Deep learning}, vol.~1.
\newblock MIT Press Cambridge, 2016.

\bibitem{grover2016node2vec}
A.~Grover and J.~Leskovec.
\newblock node2vec: Scalable feature learning for networks.
\newblock In {\em Proceedings of the 22nd ACM SIGKDD International Conference
  on Knowledge Discovery and Data Mining}, pp. 855--864. ACM, 2016.

\bibitem{hachul2004drawing}
S.~Hachul and M.~J{\"u}nger.
\newblock Drawing large graphs with a potential-field-based multilevel
  algorithm.
\newblock In {\em International Symposium on Graph Drawing}, pp. 285--295.
  Springer, 2004.

\bibitem{haleem2018evaluating}
H.~Haleem, Y.~Wang, A.~Puri, S.~Wadhwa, and H.~Qu.
\newblock Evaluating the readability of force directed graph layouts: A deep
  learning approach.
\newblock {\em arXiv preprint arXiv:1808.00703}, 2018.

\bibitem{hamilton2017inductive}
W.~Hamilton, Z.~Ying, and J.~Leskovec.
\newblock Inductive representation learning on large graphs.
\newblock In {\em Proceedings of Advances in Neural Information Processing
  Systems}, pp. 1024--1034, 2017.

\bibitem{harel2000fast}
D.~Harel and Y.~Koren.
\newblock A fast multi-scale method for drawing large graphs.
\newblock In {\em International Symposium on Graph Drawing}, pp. 183--196.
  Springer, 2000.

\bibitem{harel2002graph}
D.~Harel and Y.~Koren.
\newblock Graph drawing by high-dimensional embedding.
\newblock In {\em International Symposium on Graph Drawing}, pp. 207--219.
  Springer, 2002.

\bibitem{heimann2017generalizing}
M.~Heimann and D.~Koutra.
\newblock On generalizing neural node embedding methods to multi-network
  problems.
\newblock In {\em KDD MLG Workshop}, 2017.

\bibitem{henaff2015deep}
M.~Henaff, J.~Bruna, and Y.~LeCun.
\newblock Deep convolutional networks on graph-structured data.
\newblock {\em arXiv preprint arXiv:1506.05163}, 2015.

\bibitem{herman2000graph}
I.~Herman, G.~Melan{\c{c}}on, and M.~S. Marshall.
\newblock Graph visualization and navigation in information visualization: A
  survey.
\newblock {\em IEEE Transactions on Visualization and Computer Graphics},
  6(1):24--43, 2000.

\bibitem{hochreiter1997long}
S.~Hochreiter and J.~Schmidhuber.
\newblock Long short-term memory.
\newblock {\em Neural Computation}, 9(8):1735--1780, 1997.

\bibitem{hohman2018visual}
F.~M. Hohman, M.~Kahng, R.~Pienta, and D.~H. Chau.
\newblock Visual analytics in deep learning: An interrogative survey for the
  next frontiers.
\newblock {\em IEEE Transactions on Visualization and Computer Graphics}, 2018.

\bibitem{hu2005efficient}
Y.~Hu.
\newblock Efficient, high-quality force-directed graph drawing.
\newblock {\em Mathematica Journal}, 10(1):37--71, 2005.

\bibitem{hu2012embedding}
Y.~Hu, S.~G. Kobourov, and S.~Veeramoni.
\newblock Embedding, clustering and coloring for dynamic maps.
\newblock In {\em 2012 IEEE Pacific Visualization Symposium}, pp. 33--40. IEEE,
  2012.

\bibitem{huang2009nodeLinkEffect}
W.~Huang, P.~Eades, and S.-H. Hong.
\newblock Measuring effectiveness of graph visualizations: A cognitive load
  perspective.
\newblock {\em Information Visualization}, 8(3):139--152, 2009.

\bibitem{jacomy2014forceatlas2}
M.~Jacomy, T.~Venturini, S.~Heymann, and M.~Bastian.
\newblock Forceatlas2, a continuous graph layout algorithm for handy network
  visualization designed for the gephi software.
\newblock {\em PloS one}, 9(6):e98679, 2014.

\bibitem{kamada1989algorithm}
T.~Kamada, S.~Kawai, et~al.
\newblock An algorithm for drawing general undirected graphs.
\newblock {\em Information Processing Letters}, 31(1):7--15, 1989.

\bibitem{kaufmann2003drawing}
M.~Kaufmann and D.~Wagner.
\newblock {\em Drawing graphs: methods and models}, vol. 2025.
\newblock Springer, 2003.

\bibitem{kingma2014adam}
D.~P. Kingma and J.~Ba.
\newblock Adam: A method for stochastic optimization.
\newblock {\em arXiv preprint arXiv:1412.6980}, 2014.

\bibitem{kipf2016semi}
T.~N. Kipf and M.~Welling.
\newblock Semi-supervised classification with graph convolutional networks.
\newblock {\em arXiv preprint arXiv:1609.02907}, 2016.

\bibitem{klammler2018aesthetic}
M.~Klammler, T.~Mchedlidze, and A.~Pak.
\newblock Aesthetic discrimination of graph layouts.
\newblock In {\em International Symposium on Graph Drawing and Network
  Visualization}, pp. 169--184. Springer, 2018.

\bibitem{kruiger2017graph}
J.~F. Kruiger, P.~E. Rauber, R.~M. Martins, A.~Kerren, S.~Kobourov, and A.~C.
  Telea.
\newblock Graph layouts by t-sne.
\newblock In {\em Computer Graphics Forum}, vol.~36, pp. 283--294. Wiley Online
  Library, 2017.

\bibitem{kwon2018would}
O.-H. Kwon, T.~Crnovrsanin, and K.-L. Ma.
\newblock What would a graph look like in this layout? a machine learning
  approach to large graph visualization.
\newblock {\em IEEE Transactions on Visualization and Computer Graphics},
  24(1):478--488, 2018.

\bibitem{lancichinetti2008graphGeneration}
A.~Lancichinetti, S.~Fortunato, and F.~Radicchi.
\newblock Benchmark graphs for testing community detection algorithms.
\newblock {\em Physical Review. E, Statistical, Nonlinear, and Soft Matter
  Physics}, 78:046110, 11 2008.

\bibitem{lecun2015deep}
Y.~LeCun, Y.~Bengio, and G.~Hinton.
\newblock Deep learning.
\newblock {\em Nature}, 521(7553):436, 2015.

\bibitem{li2015gatedgraph}
Y.~Li, D.~Tarlow, M.~Brockschmidt, and R.~Zemel.
\newblock Gated graph sequence neural networks.
\newblock {\em arXiv preprint arXiv:1511.05493}, 2015.

\bibitem{masui1994evolutionary}
T.~Masui.
\newblock Evolutionary learning of graph layout constraints from examples.
\newblock In {\em Proceedings of the 7th annual ACM Symposium on User Interface
  Software and Technology}, pp. 103--108. ACM, 1994.

\bibitem{meyer1998self}
B.~Meyer.
\newblock Self-organizing graphs—a neural network perspective of graph
  layout.
\newblock In {\em International Symposium on Graph Drawing}, pp. 246--262.
  Springer, 1998.

\bibitem{monti2017geometric}
F.~Monti, D.~Boscaini, J.~Masci, E.~Rodol{\`a}, J.~Svoboda, and M.~M.
  Bronstein.
\newblock Geometric deep learning on graphs and manifolds using mixture model
  cnns.
\newblock In {\em IEEE Conference on Computer Vision and Pattern Recognition},
  pp. 5425--5434. IEEE, 2017.

\bibitem{murdoch2019interpretable}
W.~J. Murdoch, C.~Singh, K.~Kumbier, R.~Abbasi-Asl, and B.~Yu.
\newblock Interpretable machine learning: definitions, methods, and
  applications.
\newblock {\em arXiv preprint arXiv:1901.04592}, 2019.

\bibitem{niepert2016learning}
M.~Niepert, M.~Ahmed, and K.~Kutzkov.
\newblock Learning convolutional neural networks for graphs.
\newblock In {\em Proceedings of International Conference on Machine Learning},
  pp. 2014--2023, 2016.

\bibitem{noack2007energy}
A.~Noack.
\newblock Energy models for graph clustering.
\newblock {\em J. Graph Algorithms Appl.}, 11(2):453--480, 2007.

\bibitem{pascanu2013difficulty}
R.~Pascanu, T.~Mikolov, and Y.~Bengio.
\newblock On the difficulty of training recurrent neural networks.
\newblock In {\em Proceedings of International Conference on Machine Learning},
  pp. 1310--1318, 2013.

\bibitem{peng2017graphlstm}
N.~Peng, H.~Poon, C.~Quirk, K.~Toutanova, and W.-t. Yih.
\newblock Cross-sentence n-ary relation extraction with graph lstms.
\newblock {\em arXiv preprint arXiv:1708.03743}, 2017.

\bibitem{perozzi2014deepwalk}
B.~Perozzi, R.~Al-Rfou, and S.~Skiena.
\newblock Deepwalk: Online learning of social representations.
\newblock In {\em Proceedings of the 20th ACM SIGKDD International Conference
  on Knowledge Discovery and Data Mining}, pp. 701--710. ACM, 2014.

\bibitem{purchase2002metrics}
H.~C. Purchase.
\newblock Metrics for graph drawing aesthetics.
\newblock {\em Journal of Visual Languages \& Computing}, 13(5):501--516, 2002.

\bibitem{rosete1999study}
A.~Rosete-Suarez, M.~Sebag, and A.~Ochoa-Rodriguez.
\newblock A study of evolutionary graph drawing.
\newblock 1999.

\bibitem{sponemann2014evolutionary}
M.~Sp{\"o}nemann, B.~Duderstadt, and R.~von Hanxleden.
\newblock Evolutionary meta layout of graphs.
\newblock In {\em International Conference on Theory and Application of
  Diagrams}, pp. 16--30. Springer, 2014.

\bibitem{tai2015tree-lstm}
K.~S. Tai, R.~Socher, and C.~D. Manning.
\newblock Improved semantic representations from tree-structured long
  short-term memory networks.
\newblock In {\em Proceedings of the 7th International Joint Conference on
  Natural Language Processing}, vol.~1, pp. 1556--1566, 2015.

\bibitem{tamassia2013handbook}
R.~Tamassia.
\newblock {\em Handbook of graph drawing and visualization}.
\newblock Chapman and Hall/CRC, 2013.

\bibitem{tutte1960convex}
W.~T. Tutte.
\newblock Convex representations of graphs.
\newblock In {\em Proceedings of the London Mathematical Society}, vol.~3, pp.
  304--320. Wiley Online Library, 1960.

\bibitem{tutte1963draw}
W.~T. Tutte.
\newblock How to draw a graph.
\newblock {\em Proceedings of the London Mathematical Society}, 3(1):743--767,
  1963.

\bibitem{von2011visual}
T.~Von~Landesberger, A.~Kuijper, T.~Schreck, J.~Kohlhammer, J.~J. van Wijk,
  J.-D. Fekete, and D.~W. Fellner.
\newblock Visual analysis of large graphs: state-of-the-art and future research
  challenges.
\newblock {\em Computer Graphics Forum}, 30(6):1719--1749, 2011.

\bibitem{wang2016structural}
D.~Wang, P.~Cui, and W.~Zhu.
\newblock Structural deep network embedding.
\newblock In {\em Proceedings of the 22nd ACM SIGKDD International Conference
  on Knowledge Discovery and Data Mining}, pp. 1225--1234. ACM, 2016.

\bibitem{wang2005artificial}
R.-L. Wang and Okazaki.
\newblock Artificial neural network for minimum crossing number problem.
\newblock In {\em Proceedings of 2005 International Conference on Machine
  Learning and Cybernetics.}, vol.~7, pp. 4201--4204. IEEE, 2005.

\bibitem{wang2016ambiguityvis}
Y.~Wang, Q.~Shen, D.~Archambault, Z.~Zhou, M.~Zhu, S.~Yang, and H.~Qu.
\newblock Ambiguityvis: Visualization of ambiguity in graph layouts.
\newblock {\em IEEE Transactions on Visualization and Computer Graphics},
  22(1):359--368, 2016.

\bibitem{wang2017revisiting}
Y.~Wang, Y.~Wang, Y.~Sun, L.~Zhu, K.~Lu, C.-W. Fu, M.~Sedlmair, O.~Deussen, and
  B.~Chen.
\newblock Revisiting stress majorization as a unified framework for interactive
  constrained graph visualization.
\newblock {\em IEEE transactions on visualization and computer graphics},
  24(1):489--499, 2017.

\bibitem{yoghourdjian2018exploring}
V.~Yoghourdjian, D.~Archambault, S.~Diehl, T.~Dwyer, K.~Klein, H.~C. Purchase,
  and H.-Y. Wu.
\newblock Exploring the limits of complexity: A survey of empirical studies on
  graph visualisation.
\newblock {\em Visual Informatics}, 2(4):264--282, 2018.

\bibitem{you2018graphrnn}
J.~You, R.~Ying, X.~Ren, W.~L. Hamilton, and J.~Leskovec.
\newblock Graphrnn: a deep generative model for graphs.
\newblock In {\em Proceedings of the 35th International Conference on Machine
  Learning}, 2018.

\bibitem{zhang2018visual}
Q.~Zhang and S.~Zhu.
\newblock Visual interpretability for deep learning: a survey.
\newblock {\em Frontiers of Information Technology \& Electronic Engineering},
  19(1):27--39, 2018.

\bibitem{zhou2018gnnsurvey}
J.~Zhou, G.~Cui, Z.~Zhang, C.~Yang, Z.~Liu, and M.~Sun.
\newblock Graph neural networks: A review of methods and applications.
\newblock {\em arXiv preprint arXiv:1812.08434}, 2018.

\end{thebibliography}
\end{document}